\documentclass[%
 reprint,
 superscriptaddress,
 nofootinbib,
 amsmath,amssymb,
 aps,
floatfix
]{revtex4-1}
\makeatletter
\newcommand*{\rom}[1]{\expandafter\@slowromancap\romannumeral #1@}
\makeatother

\usepackage[english]{babel}
\usepackage[dvipsnames]{xcolor}
\usepackage{titlesec}
\usepackage{graphicx}
\usepackage{amsmath,amssymb,mathrsfs}
\usepackage{bigints}
\usepackage{subfig}
\usepackage{appendix}
\usepackage{graphicx}
\usepackage{dcolumn}
\usepackage{bm}
\usepackage{txfonts}
\usepackage[utf8]{inputenc}
\usepackage{grffile} 
\usepackage[justification=raggedright,singlelinecheck=off]{caption}
\usepackage{mathtools}
\usepackage{tabularx}
\usepackage{enumitem}

\newcommand{\be}{\begin{equation}}
\newcommand{\ee}{\end{equation}}
\newcommand{\ba}{\begin{eqnarray}}
\newcommand{\ea}{\end{eqnarray}}

\newcommand{\beq}{\begin{equation}}
\newcommand{\eeq}{\end{equation}}  
\newcommand{\bea}{\begin{eqnarray}}
\newcommand{\eea}{\end{eqnarray}}
\newcommand{\beqa}{\begin{eqnarray}}
\newcommand{\eeqa}{\end{eqnarray}}
\newcommand{\bseq}{\begin{subequations}}
\newcommand{\eseq}{\end{subequations}}

\newcommand{\qm}[1]{``#1''} 
\newcommand{\unitvec}[1]{\hat {{\bm #1}}} 

\definecolor{M_Green}        {rgb}{0.03 , 0.6 , 0.35}

\def\r{{\boldsymbol r}}
\def\b{{\boldsymbol b}}

\def\x{{\boldsymbol x}}

\def\k{{\boldsymbol k}}

\def\v{{\boldsymbol v}}
\def\0{{\boldsymbol 0}}
\def\v{{\boldsymbol v}}

\def\B{{\boldsymbol B}}

\def\E{{\boldsymbol E}}

\def\nab{{\boldsymbol \nabla}}

\def\cal{\mathcal}

\def\detmet{\vert g \vert^{\frac{1}{2}}}

\begin{document}

\title{Numerical magneto-hydrodynamics for relativistic nuclear collisions} 

\author{Gabriele~Inghirami}
\email{inghirami@fias.uni-frankfurt.de}
\affiliation{Frankfurt Institute for Advanced Studies, Ruth-Moufang-Straße 1,
 60438 Frankfurt am Main, Germany}
\affiliation{Institute for Theoretical Physics, Goethe-Universit{\"a}t, Max-von-Laue-Straße 1,
 60438 Frankfurt am Main, Germany}
\affiliation{GSI Helmholtzzentrum für Schwerionenforschung GmbH,
Planckstraße 1,
64291 Darmstadt, Germany}
\affiliation{John von Neumann Institute for Computing,
Forschungszentrum Jülich,
52425 Jülich, Germany}

\author{Luca~\surname{Del~Zanna}}
\affiliation{Dipartimento di Fisica e Astronomia, Universit\`a di Firenze, 
Via G. Sansone 1, I-50019 Sesto F.no (Firenze), Italy}
\affiliation{INFN - Sezione di Firenze, Via G. Sansone 1, I-50019 Sesto F.no (Firenze), Italy}
\affiliation{INAF - Osservatorio Astrofisico di Arcetri, L.go E. Fermi 5, I-50125 Firenze, Italy}

\author{Andrea~Beraudo}
\affiliation{INFN - Sezione di Torino, Via P. Giuria 1, I-10125 Torino, Italy}

\author{Mohsen~\surname{Haddadi Moghaddam}}
\affiliation{Department of Physics, Hakim Sabzevari University, P. O. Box 397, Sabzevar, Iran}
\affiliation{INFN - Sezione di Torino, Via P. Giuria 1, I-10125 Torino, Italy}

\author{Francesco~Becattini}
\affiliation{Dipartimento di Fisica e Astronomia, Universit\`a di Firenze, 
Via G. Sansone 1, I-50019 Sesto F.no (Firenze), Italy}
\affiliation{INFN - Sezione di Firenze, Via G. Sansone 1, I-50019 Sesto F.no (Firenze), Italy}

\author{Marcus~Bleicher}
\affiliation{Frankfurt Institute for Advanced Studies, Ruth-Moufang-Straße 1,
 60438 Frankfurt am Main, Germany}
\affiliation{Institute for Theoretical Physics, Goethe-Universit{\"a}t, Max-von-Laue-Straße 1,
 60438 Frankfurt am Main, Germany}
 \affiliation{GSI Helmholtzzentrum für Schwerionenforschung GmbH,
Planckstraße 1,
64291 Darmstadt, Germany}
\affiliation{John von Neumann Institute for Computing,
Forschungszentrum Jülich,
52425 Jülich, Germany}
 
\date{\today}
\begin{abstract}We present an improved version of the ECHO-QGP numerical code, which self-consistently includes for the first time the effects of electromagnetic fields within the framework of relativistic magnetohydrodynamics (RMHD). We discuss results of its application in relativistic heavy-ion collisions in the limit of infinite electrical conductivity of the plasma. After reviewing the relevant covariant $3\!+\!1$ formalisms, we illustrate the implementation of the evolution equations in the code and show the results of several tests aimed at assessing the accuracy and robustness of the implementation. After providing some estimates of the magnetic fields arising in non-central high-energy nuclear collisions, we perform full RMHD simulations of the evolution of the Quark-Gluon Plasma in the presence of electromagnetic fields and discuss the results. In our ideal RMHD setup we find that the magnetic field developing in non-central collisions does not significantly modify the elliptic-flow of the final hadrons. However, since there are uncertainties in the description of the pre-equilibrium phase and also in the properties of the medium, a more extensive survey of the possible initial conditions as well as the inclusion of dissipative effects are indeed necessary to validate this preliminary result.  
\end{abstract}


\maketitle

\section{Introduction}
High-energy nuclear collisions, studied by several experimental collaborations at RHIC and at the LHC, allow one to explore the QCD phase-diagram in the high-temperature region, from high to almost vanishing baryonic density. Strong evidence, coming both from soft and hard observables, was obtained for the onset of a deconfined phase in the RHIC and LHC energy regime. Furthermore, at the experimentally accessible conditions (i.e. slightly above the deconfinement phase-transition), the produced system, with a lifetime $\sim$ 10 fm/c, was found to behave like a collective, strongly-interacting medium, rather opaque to penetrating probes, in contrast to the expected gas of weakly-interacting quarks and gluons. Relativistic hydrodynamic models (nowadays including also dissipative effects) were developed to describe the evolution -- driven by pressure gradients -- of the produced matter and turned out to reproduce the data quite well~\cite{PhysRevLett.99.172301,PhysRevC.77.064901,PhysRevC.85.034901,PhysRevLett.110.012302,PhysRevLett.97.202302,DelZanna:2013eua,Becattini:2015ska,Karpenko20143016}, in particular the various flow-harmonics arising from the collective response of the system to the anisotropies and fluctuations in the initial conditions.

While the main purpose of relativistic heavy-ion experiments is the study of strong interactions at extreme energy densities similar to the early universe, it was recently realized that during the collisions of high-$Z$ nuclei ($Z\!=\!82$ for Pb) at ultra-relativistic energies, one can also produce the strongest magnetic fields reached in our universe, with initial values of $B\!\sim\!10^{15}$ T and oriented mainly in the direction perpendicular to the reaction-plane~\cite{KHARZEEV2008227}.
In the last years it was suggested~\cite{KHARZEEV2008227,PhysRevD.78.074033} that, besides leading to the production of a strongly-interacting deconfined system, the presence of these strong magnetic fields in relativistic heavy-ion collisions opens also the possibility of exploring peculiar non-perturbative features of QCD, such as the appearance of non-trivial topological configurations of the color-field. Once coupled to quarks, these configurations characterized by a non-vanishing winding number lead to an excess of quarks of a given chirality (chiral anomaly), depending on the value of the topological charge, and hence, on an event-by-event basis, to a violation of parity (clearly preserved after an event-average). In the presence of strong magnetic fields this can give rise to observable effects, with a separation of oppositely-charged particles with respect to the reaction-plane. Since for massless particles with a fixed handedness (e.g. right handed quarks) the chirality coincides with the helicity (i.e. the projection of the spin along the particle momentum) and since particles tend to align their magnetic moments along the $B$-field, one would have an excess of positively-charged $u$-quarks moving in the direction of the magnetic field and an excess of negative $d$-quarks moving in the opposite direction. Clearly, averaging over a large sample of events, each one with a different excess of right or left-handed quarks, the effect should cancel at the level of single-particle distributions; however, it should leave its fingerprints in multi-particle correlations, as suggested in~\cite{PhysRevC.70.057901}. Due to the interplay between a non-perturbative feature of strong interactions (the chiral anomaly) and the role of the magnetic field, such a phenomenon was called Chiral Magnetic Effect (CME) and is currently studied by different experimental collaborations at RHIC and at the LHC~\cite{PhysRevC.81.054908,PhysRevLett.103.251601,PhysRevC.93.044903}. Analogous effects have been recently observed also in astrophysics (as an explanation of Neutron Stars kicks)~\cite{Kaminski2016170} and in solid-state physics, placing Dirac semi-metals in parallel magnetic and electric fields~\cite{Li:2014bha,Xiong:2015nna,Huang:2015eia,arnold-shekar16}. Other related  phenomena (Chiral Magnetic Wave~\cite{PhysRevD.83.085007}, Chiral Separation Effect~\cite{PhysRevD.22.3067}, Chiral Vortical Effect~\cite{Kharzeev20161}), all arising from an unbalance among right and left-handed particles and from the presence of a strong magnetic field or angular momentum, were suggested to occur in non-central heavy-ion collisions: for an overview we refer the reader to~\cite{Kharzeev20161}.    

An unambiguous observation of the CME in heavy-ion collisions would be clearly a result of deep theoretical interest, since it would represent a manifestation of the non-trivial topological structure of a Yang-Mills theory. However, in order to separate opposite-sign charges with respect to the reaction-plane, the initial magnetic field generated by the colliding nuclei must be sufficiently long lived. The lifetime of the magnetic field depends strongly on the nature of the produced medium. In the vacuum the initial magnetic field decays rather rapidly. On the contrary in the opposite limit, in the presence of an ideal plasma with infinite electric conductivity, the freezing of the magnetic-flux makes the field survive much longer and may allow for the manifestation of signatures of the possible chiral unbalance in the final charged-hadron spectra, even though, at the same time, a large conductivity would also tend to compensate any local charge excess. 
Unfortunately, so far in the literature one can find only semi-analytic estimates of the time-evolution of the magnetic field in heavy-ion collisions, based on simplifying assumptions~\cite{Tuchin:2013apa,Tuchin:2013ie,McLerran:2013hla,Deng:2012pc,Gursoy:2014aka,Bzdak:2011yy}. A fully realistic calculation would require to solve the the Maxwell equations together with the continuity equations for the energy-momentum tensor (closed by some form of Ohm's law), i.e. it calls for a full Relativistic Magneto-HydroDynamic (RMHD) description of the medium, in which the evolution of the electromagnetic field is consistently coupled with the evolution of the plasma: this is the challenge we address with the present paper. 

For this first study we consider the case of an ideal plasma, with no dissipative effects and, in particular, an infinite electric conductivity, which makes the electric field in the local rest-frame of the medium vanish. We also neglected any anomalous term in the currents, although previous studies \cite{Son:2009tf,Kharzeev:2011ds} in simplified models showed that they would not to contribute to entropy production, being in this sense ``ideal'': the inclusion of dissipative and anomalous terms (necessary for the description of the CME) in our setup is left for future work. In light of the small experimental uncertainties reached at the LHC and RHIC on flow measurements the development of a code able to consistently treat the coupled evolution of the plasma and $Z$-enhanced electromagnetic fields represents in any case a necessary baseline for any claim that CME (and other related phenomena that we will be able to address after including anomalous currents) can be disentangled from possible other confounding electromagnetic effects that could lead to charge separation.

Our paper is organized as follows. In Sec.~\ref{sec:setup} we present the RMHD equations in their most general form, focusing then on their ideal limit, i.e. on the case of a plasma with infinite electrical conductivity (and neglecting other dissipative effects such as viscosity and thermal conduction). Only the ideal case is considered for the present paper. In Sec.~\ref{sec:implementation} we discuss the numerical implementation of the ideal-RMHD equations, written in a conservative form, within our improved ECHO-QGP code. In Sec.~\ref{sec:tests} we discuss the results of a large variety of numerical tests to prove the accuracy and the robustness of the implementation: the shock-tube problem, the description of Alfv\'en waves, the rotor test, the reproduction of the one-dimensional Bjorken expansion in a magnetic field and the accurate treatment of the in-vacuum self-similar expansion in transverse-MHD.
In Sec.~\ref{sec:results} we show the results obtained from the code with simplified (but reasonable) initial conditions for non-central nucleus-nucleus collisions. At least in the context of this simplified approach, the magnetic field is not able to modify the elliptic flow of the final hadrons substantially. Nevertheless, further and more realistic investigations are needed before solid conclusions can be drawn. Finally, in Sec.~\ref{sec:conclusions} we discuss our findings and the future perspectives of our work, with the idea of performing 3D+1 simulations based on a much broader pool of different initial conditions, possibly including dissipative effects.
The appendix is devoted to a discussion of the propagation of linear perturbations in RMHD, focusing on the case of fast-magnetosonic and Alfv\'en waves, which are the ones relevant for the analysis carried out in this paper.

\section{Ideal relativistic magnetohydrodynamics}\label{sec:setup}

Relativistic MHD (RMHD hereafter) is a one-fluid description of the interaction of matter and electromagnetic fields in plasmas \cite{bekenstein78,an89}. In general, as in the Newtonian limit of classical MHD, one assumes that there is a dominant species determining a main \emph{fluid} current, while a secondary species must be responsible for the conduction current, namely the source for the electromagnetic field. The RMHD evolution equations describing the dynamics of the overall system are the conservation laws for this fluid current $N^\mu$ (associated to the net-baryon current or to any other conserved charge, if any) and for the \emph{total} (matter and fields) energy-momentum tensor of the plasma $T^{\mu\nu}$, namely
\begin{align}
& d_\mu N^\mu  =  0, \label{eq:mass} \\
& d_\mu T^{\mu\nu}  =  0 \label{eq:enmom} ,
\end{align}
with $d_{\mu}$ being the covariant derivative,
thus to be supplemented by the second law of thermodynamics 
\be
d_\mu s^\mu \ge 0,
\label{eq:entropy}
\ee
where $s^\mu$ is the entropy current. On the other hand, the electromagnetic field obeys Maxwell's equations
\begin{align}
& d_\mu F^{\mu\nu} = - J^\nu \quad (d_\mu J^\mu = 0),  \label{eq:maxwell1} \\
& d_\mu F^{\star\mu\nu} = 0,     \label{eq:maxwell2}
\end{align}
where $F^{\mu\nu}$ is the Faraday tensor and $F^{\star\mu\nu}=\textstyle{\frac{1}{2}}\epsilon^{\mu\nu\lambda\kappa}F_{\lambda\kappa}$ is its dual. Notice that here we have neglected possible polarization and magnetization effects of the plasma, therefore we do not make a distinction between microscopic and macroscopic fields~\cite{dixon1978}. Under this assumption, the electromagnetic contribution to the energy-momentum tensor is known to be
\be
T^{\mu\nu} _\mathrm{f} = F^{\mu\lambda}F^\nu_{\,\lambda} - 
\tfrac{1}{4}g^{\mu\nu}F^{\lambda\kappa}F_{\lambda\kappa},
\ee
for which $d_\mu T^{\mu\nu} _\mathrm{f}=J_\mu\,F^{\mu\nu}$, from Maxwell equations. Introducing the matter contribution to the energy-momentum tensor $T^{\mu\nu} _\mathrm{m}$ and letting $T^{\mu\nu} = T^{\mu\nu} _\mathrm{m} + T^{\mu\nu} _\mathrm{f}$, Eq.~(\ref{eq:enmom}) gives
\be
d_\mu T^{\mu\nu} _\mathrm{m} = - J_\mu F^{\mu\nu},
\label{eq:dTm}
\ee
where the right-hand-side is the Lorentz force acting on the plasma.

In the ideal limit all dissipative fluxes can be neglected and local equilibrium is assumed. A single \emph{fluid four-velocity} $u^\mu$ ($u_\mu u^\mu = -1$) can be thus defined and we write
\begin{align}
& N^\mu = n u^\mu, \\
& T^{\mu\nu}_\mathrm{m} =
e u^\mu u^\nu + p \Delta^{\mu\nu} = (e + p) u^\mu u^\nu + p g^{\mu\nu}, \label{eq:Tm} \\
& s^\mu  =  s u^\mu, 
\end{align}
where we have introduced the projector $\Delta^{\mu\nu} = g^{\mu\nu} + u^\mu u^\nu$ ($\Delta^{\mu\nu} u_\nu = 0$). In the above zeroth-order relations $n= - N^\mu u_\mu$ is the main charge density, $e = T^{\mu\nu}_\mathrm{m}u_\mu u_\nu $ the fluid energy density, and $p = \tfrac{1}{3} \Delta_{\mu\nu} T^{\mu\nu}_\mathrm{m}$ the kinetic pressure, all quantities are defined in the comoving frame. The Faraday tensor and its dual can also be split with respect to $u^\mu$ as
\begin{align}
F^{\mu\nu} & = u^\mu e^\nu - u^\nu e^\mu + \epsilon^{\mu\nu\lambda\kappa} b_\lambda u_\kappa, \\
F^{\star\mu\nu} & = u^\mu b^\nu - u^\nu b^\mu - 
\epsilon^{\mu\nu\lambda\kappa} e_\lambda u_\kappa,
\end{align}
where
\begin{align}
e^\mu  = F^{\mu\nu}u_\nu,  \quad & (e^\mu u_\mu =0), \\
b^\mu  = F^{\star\mu\nu}u_\nu, \quad &  (b^\mu u_\mu =0),
\end{align}
are the electric and magnetic fields measured in the comoving frame of the fluid. 

Since the electromagnetic fields do not evolve in vacuum, but are strongly coupled with the fluid, we must now provide an appropriate Ohm law relating the current with the fields. In the simplest case one usually assumes the linear form
\be
J^\mu = \rho_\mathrm{e} u^\mu + j^{\,\mu}; \quad j^{\,\mu} = \sigma^{\,\mu\nu} e_\nu,
\ee
where $\rho_\mathrm{e}$ is the electric charge density in the comoving frame, $j^{\,\mu}$ the conduction current ($j^{\,\mu} u_\mu = 0$), and $\sigma^{\,\mu\nu}$ the plasma conductivity tensor. The presence of a finite conductivity in the plasma gives rise to (anisotropic) magnetic dissipation and Joule heating, as well as to topological field line changes known as \emph{magnetic reconnection}. Recent theoretical and numerical results may be found in \cite{DelZanna21082016} and references therein. 

In the ideal MHD approximation considered in the present paper we assume a conductivity high enough to avoid the onset of huge currents in the plasma. We can then replace the Ohm law with its limiting case
\be
e^\mu = 0.
\label{eq:ohm}
\ee
When the above condition holds, the expressions for the Faraday tensor and for its dual are simplified, and the number of unknowns is reduced. In particular,  Eq.~(\ref{eq:maxwell1}) will be used to derive the current, if needed, while Eq.~(\ref{eq:maxwell2}) will become the evolution equation for $b^\mu$. Moreover, the electromagnetic energy-momentum tensor becomes
\begin{align}
T^{\mu\nu} _\mathrm{f} & = \tfrac{1}{2} b^2 u^\mu u^\nu +  \tfrac{1}{2} b^2 \Delta^{\mu\nu} - b^\mu b^\nu \nonumber \\
& = b^2 u^\mu u^\nu +  \tfrac{1}{2} b^2 g^{\mu\nu} - b^\mu b^\nu,
\end{align}
where $b^2 = b_\mu b^\mu$, which can be plugged into Eq.~(\ref{eq:enmom}) together with the corresponding matter contribution in Eq.~(\ref{eq:Tm}). Summarizing, the system of ideal RMHD equations is
\begin{align}
& d_\mu ( n u^\mu) = 0,  \label{eq:mhd1} \\
& d_\mu [(e + p + b^2) u^\mu u^\nu +  ( p + \tfrac{1}{2} b^2) g^{\mu\nu} - b^\mu b^\nu] = 0,  \label{eq:mhd2} \\
& d_\mu ( u^\mu b^\nu - u^\nu b^\mu ) = 0  \label{eq:mhd3} , 
\end{align}
in the unknowns $n$, $e$, $p$, $u^\mu$, and $b^\mu$. 

Non-conservative versions of the above equations can also be found. It is useful to decompose the covariant derivative as
\be
d_\mu = - u_\mu D + \nabla_\mu,
\ee
where $D\!\equiv\!u^\mu d_\mu$ indicates derivation along $u^\mu$ (reducing to the Eulerian time derivative in the nonrelativistic limit), and $\nabla_\mu = \Delta_\mu^\nu d_\nu$ is the derivative transverse to the flow (reducing to the spatial gradient in the nonrelativistic limit). The charge conservation (baryon-number in the case of heavy-ion collisions) becomes
\be
Dn + n \theta = 0, \label{eq:mass2}
\ee
where $\theta\!\equiv\! d_\mu u^\mu\! = \!\nabla_\mu u^\mu$ is the \emph{expansion factor}. The energy equation is derived by projecting the $d_\mu T^{\mu\nu}\!=\!0$ conservation-law along the flow $u_\nu$, where, we remember, the total energy-momentum tensor is given by the sum of the matter and field components: $T^{\mu\nu}=T_{\rm m}^{\mu\nu}+T_{\rm f}^{\mu\nu}$. From Eq.~(\ref{eq:dTm}) we get
\be
u_\nu d_\mu T^{\mu\nu}_\mathrm{m} = - J_\mu F^{\mu\nu}u_\nu,
\ee
which leads to
\beq
D e + (e+p) \theta =J_\mu e^\mu.
\eeq
Written in the above form, the energy equation is rather general, the right-hand side representing the Joule heating of the fluid.
However, as previously discussed, in ideal MHD the electric field in the local rest-frame vanishes, $e^\mu\!=\!0$, thus one simply has
\be
D e + (e+p) \theta = 0, \label{eq:energy2}
\ee
independent of $b^\mu$, as in ordinary relativistic hydrodynamics. This form of the energy equation will be exploited in discussing the Bjorken-flow of a magnetized plasma in Sec.~\ref{sec:Bjorken}. However, if the two contributions are kept together, we may also write
\be
D (e + \tfrac{1}{2}b^2) + (e+p+b^2) \theta + u_\mu b^\nu d_\nu b^\mu= 0. \label{eq:energy3}
\ee
The relativistic extension of the MHD Euler equation is retrieved by projecting the total energy-momentum conservation law transverse to the flow, that is
\be
\begin{aligned}
	&(e+p+b^2) D u^\mu + \nabla^\mu (p + \tfrac{1}{2}b^2) \! =\\
	&b^\mu d_\nu b^\nu + b^\nu d_\nu b^\mu + u^\mu u_\nu b^\lambda d_\lambda b^\nu .
\end{aligned}
\ee
Several expressions may be derived from the last RMHD equation for the evolution of $b^\mu$, here we choose to rewrite it as
\be
D b^\mu + \theta b^\mu = u^\mu b^\nu D u_\nu + b^\nu d_\nu u^\mu,
\ee
where we have used the relation $d_\mu b^\mu = b^\mu D u_\mu$.

Finally, the system of ideal RMHD equations must be closed by choosing an equation of state (EoS), for instance of the form $p=\mathcal{P}(e, n)$, under the assumption that in the ideal case each local equilibrium state can be completely determined by $u^\mu$ and two thermodynamical variables ($e$ and $n$ in this case). The Euler and Gibbs-Duhem relations read
\be
e + p = Ts +  \mu n, \quad d e = T d s +\mu d n,
\label{eq:gibbs}
\ee
where we have defined the local temperature $T = (\partial e / \partial s)_n$ and the chemical potential $\mu = (\partial e / \partial n)_s$. Eqs~(\ref{eq:gibbs}), (\ref{eq:energy2}), and (\ref{eq:mass2}) allow us to write
\be
Ds + s \, \theta = 0.\label{eq:entro}
\ee
We then retrieve the expected result that in the ideal case, when all dissipative terms are neglected, there is no entropy production and Eq.~(\ref{eq:entropy}) holds as an equality. Notice that the entropy current is conserved even in the case of vanishing charge (baryon-number) density and chemical potential $n\!=\!\mu\!=\!0$, as appropriate for high-energy heavy-ion collisions and an ultrarelativistic EoS with $p=\mathcal {P}(e)$.

\section{The RMHD module in ECHO-QGP}\label{sec:implementation}

We now rewrite the evolution equations for ideal RMHD in a form suitable for numerical integration, for which we need a clear separation between time and space components (the so-called $3+1$ split) and the preservation of the original conservative character of the equations, since shock-capturing numerical codes such as ECHO-QGP require to solve a series of balance laws. Here we will provide the basic expressions, for further formal and technical details details see~\cite{ldz03,ldz07,DelZanna:2013eua} and references therein.

Neglecting curvature effects due to gravitational fields, we consider here a metric in special relativity (though not necessarily Minkowskian) of the form
\be
ds^2 = - dx^0 dx^0 + g_{ij}\,dx^i dx^j
\ee
where the three-metric coefficients $g_{ij}$ may depend both on space $x^i$ and time $x^0$, in general. It is first useful to introduce the fluid velocity $v^i$ and electric and magnetic fields $E^i$ and $B^i$ as measured in the laboratory frame, which are spatial vectors (vanishing time component). The fluid four velocity can be expressed as
\be
u^\mu = ( \gamma , \gamma v^i ), 
\ee
where $\gamma = (1-v^2)^{-1/2}$ is the Lorentz factor of the bulk flow and $v^2=v_kv^k$, whereas the fields are, respectively
\be
e^\mu = ( \gamma v_kE^k, \gamma E^i + \gamma \varepsilon^{ijk} v_jB_k),
\ee
\be
b^\mu = ( \gamma v_kB^k, \gamma B^i - \gamma \varepsilon^{ijk} v_jE_k),
\ee
where $\varepsilon_{ijk}$ is the Levi-Civita pseudo-tensor of the spatial three-metric, namely  $\varepsilon_{ijk} = \vert g \vert^{\frac{1}{2}} [ijk]$, with $g=\mathrm{det}\{g_{\mu\nu}\}=-\mathrm{det}\{g_{ij}\} < 0$ and $[ijk]$ the usual alternating symbol of three-dimensional space with values $\pm 1$ or 0. From the ideal Ohm law of Eq.~(\ref{eq:ohm}) we can derive the spatial electric field as
\be
E_i = - \varepsilon_{ijk}v^jB^k, \label{eq:ohm2}
\ee
which is known once $v^i$ and $B^i$ have been determined. In this case the $b^\mu$ field is
\be
b^\mu = (\gamma v_kB^k,  B^i/\gamma + \gamma v_kB^k v^i )\label{eq:bvsB_id}
\ee
with
\be
b^2 = B^2 - E^2 = B^2/\gamma^2 + (v_kB^k)^2
\ee
where $B^2=B_kB^k$ and $E^2=E_kE^k=v^2B^2 - (v_kB^k)^2$. Notice that when $v^i=0$, that is in the fluid rest frame, we retrieve $u^\mu = (1,0)$ and $b^\mu = (0, B^i)$, as expected. 

Let us now rewrite Eqs.~(\ref{eq:mhd1}-\ref{eq:mhd3}) in a form appropriate for numerical integration, by clearly separating time and space derivatives and tensor components. We find the system
\be
\label{eq:echo}
\partial_0 {\bf U}+ \partial_i {\bf F}^i ={\bf S}, 
\ee
where
\be
{\bf U} \! = \! \detmet \! \left(\begin{array}{c}
	\gamma n \\
	S_j \equiv T^0_{\,j}\\
	\mathcal{E} \equiv - T^0_{\,0} \\
	B^j
\end{array}\right), \,
{\bf F}^i \! = \! \detmet \! \left(\begin{array}{c}
	\gamma n v^i \\
	T^i_{\,j}\\
	S^i \equiv - T^i_{\, 0} \\
	v^iB^j - B^iv^j
\end{array}\right)
\label{eq:fluxes}
\ee
are respectively the set of \emph{conservative variables} and \emph{fluxes}, 
while the source terms are given by
\be
{\bf S}=\detmet \left(\begin{array}{c}
	0 \\
	\tfrac{1}{2} T^{ik}\partial_j g_{ik} \\
	- \tfrac{1}{2} T^{ik}\partial_0 g_{ik} \\
	0
\end{array}\right),
\label{eq:sources}
\ee
where the symmetric and antisymmetric properties of $T^{\mu\nu}$ and $F^{\star\mu\nu}$, respectively, have been exploited in deriving the above balance laws.

The components of $T^{\mu\nu}$ appearing in the expressions for the conserved variables and fluxes are
\begin{align}
S_i = & (e+p)\gamma^2v_i  + \varepsilon_{ijk}E^jB^k , \label{eq:S} \\
T_{ij} = & (e+p)\gamma^2 v_iv_j + (p + u_\mathrm{em} )g_{ij} -E_iE_j - B_iB_j ,  \\
\mathcal{E} = & (e+p)\gamma^2 - p + u_\mathrm{em} \label{eq:E}, 
\end{align}
where we have defined the electromagnetic energy density $u_\mathrm{em}=\tfrac{1}{2}(E^2+B^2)$. We recall that while $B^i$ is a dynamical variable, $E^i$ is a derived quantity, obtained from Eq.~(\ref{eq:ohm2}).

One final constraint comes from the time component of Eq.~(\ref{eq:mhd3}), that is the solenoidal condition
\be
\partial_i (\vert g \vert^{\frac{1}{2}} B^i) = 0,
\label{eq:solenoidal}
\ee
which, if valid at the initial time of the evolution, should be preserved analytically by the last equation of the above RMHD system. From a numerical point of view, however, this constraint needs some specific techniques to be actually enforced. In fact, the accumulation of the numerical errors associated to the computation of the derivatives of the magnetic field may lead to the violation of the solenoidal (i.e. ``null-B divergence'') condition (\ref{eq:solenoidal}), implying the formation of unphysical magnetic monopoles and fictitious forces. There are several methods to avoid, or at least to limit, this issue~\cite{evans88,powell99,lond00,lond04}. We adopted the method proposed by Dedner for MHD and later extended to the cases of special and general relativity~\citep{dedner02,Palenzuela21042009,Mignone20102117,penner11,moesta14,PhysRevD.88.044020}.

\subsection{Numerical procedures}
ECHO-QGP is based on finite difference schemes. At the beginning of the simulation, the initial values of the \emph{primitive variables} $n$ (the baryon density), $v^i$ (the contravariant components of the velocity of the fluid in the lab frame), $p$ (the pressure of the fluid in the comoving frame) and $B^i$ (the contravariant components of the magnetic induction field in the lab frame) are discretized on the computational grid by evaluating them at the center of each cell.
Time integration of conservative variables is performed using a second or third order Runge-Kutta algorithm, then, at each sub-timestep:
\begin{itemize}
	\item the values of the primitive variables are reconstructed at cell borders, for each direction (several algorithms are implemented and can be selected \cite{ldz07}: TVD2, CENO3, WENO3, WENO5, PPM4, MPE3, MPE5, MPE7),
	\item fluxes in Eq.~(\ref{eq:fluxes}) are computed,
	\item the Riemann problem for fluxes at cell interfaces is solved using the HLL (Harten-Lax-Van Leer)~\cite{hll83} approximate method,
	\item the divergence of these \emph{numerical fluxes} and source terms in Eq.~(\ref{eq:sources}) are computed at cell centers, allowing to integrate the discretized evolution equations for the conservative variables,
	\item the new primitive variables are retrieved from the evolved conservative ones.
\end{itemize}

This last step above implies to solve a system of non-linear equations and currently there is no known algorithm which guarantees a global convergence to the solutions. The system is more easily solved by providing an initial guess for the solution, usually chosen as the values of the primitive variable at the previous timestep. However, in a rapidly evolving system as in the case of heavy ion collision, this guess may not be close enough to the real solution and the algorithm may fail or converge to other (unphysical) solutions.
Nevertheless, if we restrict to the use of a specific analytic Equation of State (EoS), then the system of non linear equations may be considerably simplified and it is possible to develop very robust inversion routines~\cite{ldz07,mignone07}.

For the present study we focus for sake of simplicity on the ultra-relativistic gas EoS $p=e/3$, using an \qm{ad hoc} version of the method described in \cite{ldz07}, hereafter shortly summarized. We exploit Eq.~(\ref{eq:ohm2}) to rewrite equations (\ref{eq:S}) and (\ref{eq:E}), then we compute $S^2=S_i S^i$ and $S_i B^i$, which are known since $B^i$ is both a conservative and primitive variable (the difference is  only in the factor $\vert g \vert^{\frac{1}{2}}$). After introducing the new variables $x=v^2=v^i v_i$ and $y=4p\gamma^2$, with some algebraic manipulations we can formulate the following system of equations:
\beq
(y+B^2)^2x-y^{-2}(S_i B^i)^2(2y+B^2)-S^2=0,
\label{eq:F1}
\eeq
\beq
\dfrac{3+x}{4}y +\dfrac{1}{2}(1+x)B^2-\dfrac{1}{2}y^{-2}(S_i B^i)^2-\mathcal{E}=0.
\label{eq:F2}
\eeq
These coupled non-linear equations are solved through a nested procedure: Eq.~(\ref{eq:F1}) is solved for $x$ with a one dimensional iterative hybrid Newton-Raphson/bisection method \cite{nr} with bracketing between 0 and 1; at each iteration of this routine, the $y$ variable is obtained by finding the (unique) positive root of the third order polynomial of Eq.~(\ref{eq:F2}) multiplied by $y^2$ with $x=x(y)$.
The solution of the system allows then to compute the primitive variables through the relations:
\beq
v^i=\frac{S^i+ (S_kB^k)B^i/y}{y+B^2}, \quad 
p= \frac{e}{3} = \frac{1}{4} (1-x) y.
\eeq

For EoS where the pressure $p$ depends also on the baryon density $n$, like the ideal gas EoS used in \cite{ldz07} and in the shock tube test presented here, note that the latter quantity can be easily obtained by dividing the corresponding conserved variable by the Lorentz factor $\gamma$. However, for a comparison to high energy HIC data, a lattice QCD based equation of state should be employed~\cite{laine06} (in contrast to the simplified EoS used for the present study), which unfortunately does not allow to simplify the system of non linear equations on which the inversion routine is based and needs a more careful (and slower) numerical treatment as discussed above.

\section{Tests}\label{sec:tests}

In this section we present some numerical test problems selected in order to validate the code. We avoid to repeat tests aimed at simply measuring the accuracy of the ``core'' algorithms, since ECHO-QGP for relativistic hydrodynamics~\cite{DelZanna:2013eua,Becattini:2015ska} has been already validated against basic benchmarks, and many additional tests have been performed on the original ECHO code~\cite{ldz07}, from which ECHO-QGP has been derived sharing the same base structure. Instead, here we focus on checking the correctness of its results in the ideal RMHD context. We use the ultrarelativistic EoS $p\!=\!e/3$, if not mentioned otherwise.

We will use either Minkowski $(t,x,y,z)$ or Milne $[\tau,x,y,\eta_s]$ coordinates, where $\tau\!\equiv\!\sqrt{t^2-z^2}$ is the longitudinal proper-time and $\eta_s\equiv\frac{1}{2}\ln\frac{t+z}{t-z}$ the space-time rapidity. In the following, in writing four-vector components in Milne coordinates, we will employ square brackets. Notice that in the first case the three-metric is $g_{ij}=\mathrm{diag}\{1,1,1\}$, with $\detmet = 1$, whereas for Milne coordinates $g_{ij}=\mathrm{diag}\{1,1,\tau^2\}$, with  $\detmet = \tau$. In both cases $\partial_j g_{ik}=0$ and the source terms in the evolution equations simplify considerably. Notice that in Milne coordinates, where $g_{33}=\tau^2$, the source term for the energy equation contains a non-vanishing term proportional to $\tfrac{1}{2}\partial_0 g_{33} = \tau$.

\subsection{Magnetized shock tube}

\begin{figure*}
	\centerline{
		\includegraphics[width=0.5\textwidth]{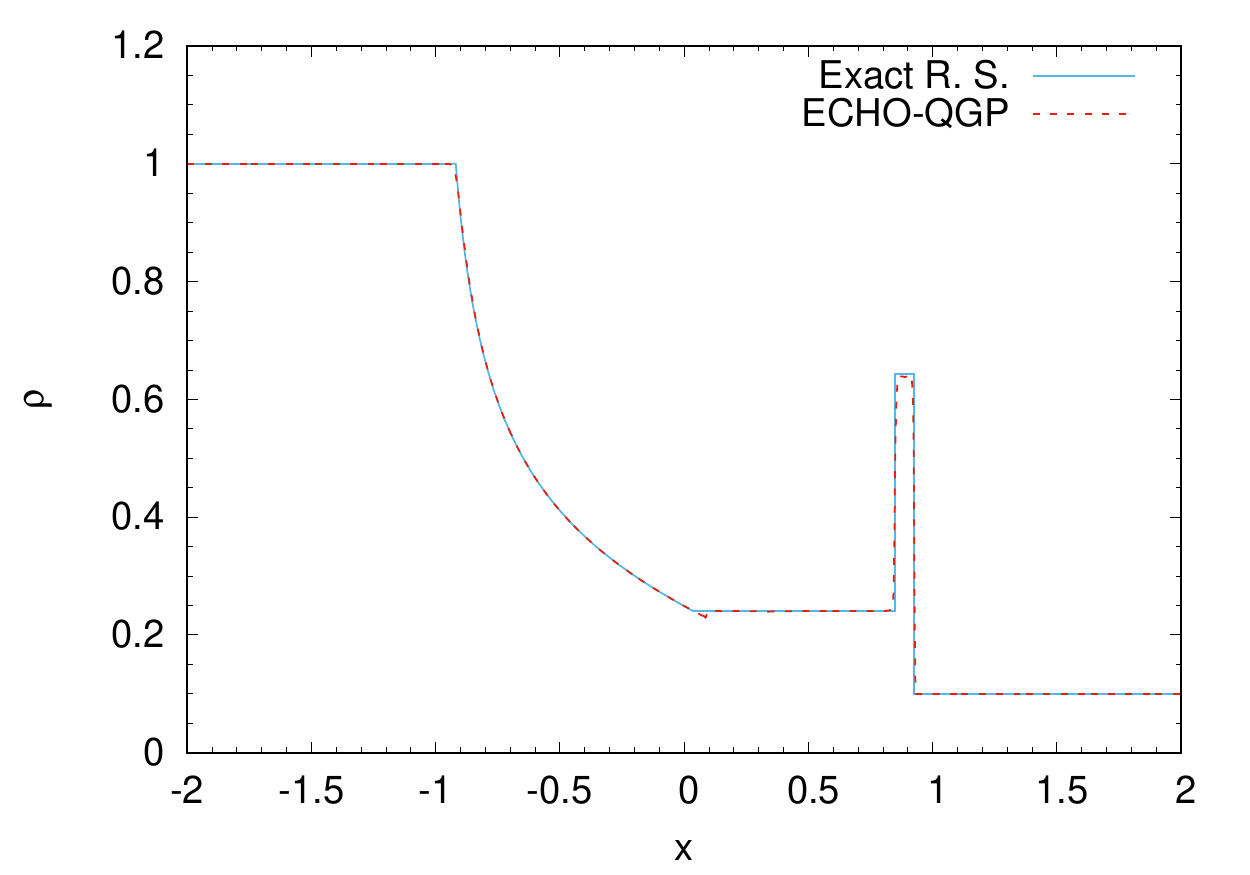}
		\includegraphics[width=0.5\textwidth]{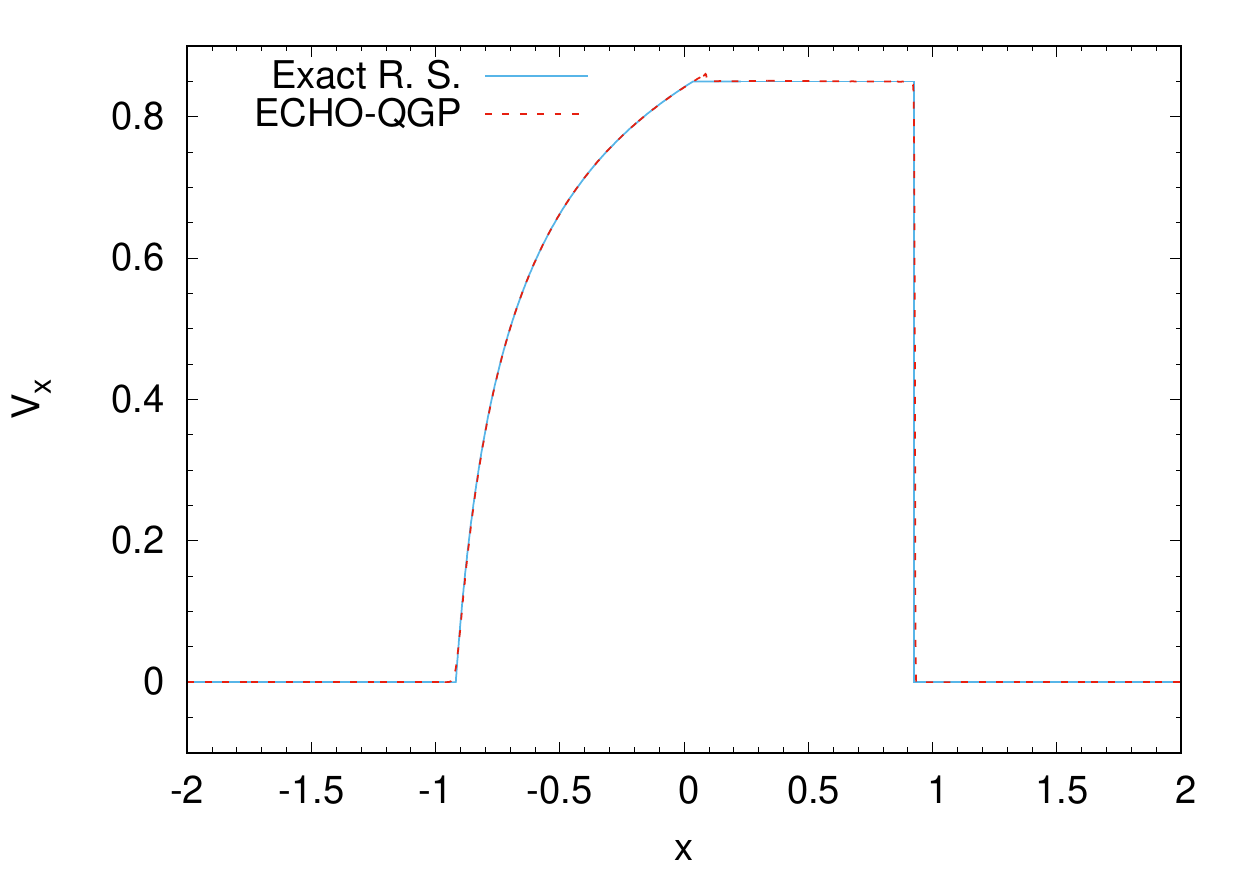}}
	\centerline{
		\includegraphics[width=0.5\textwidth]{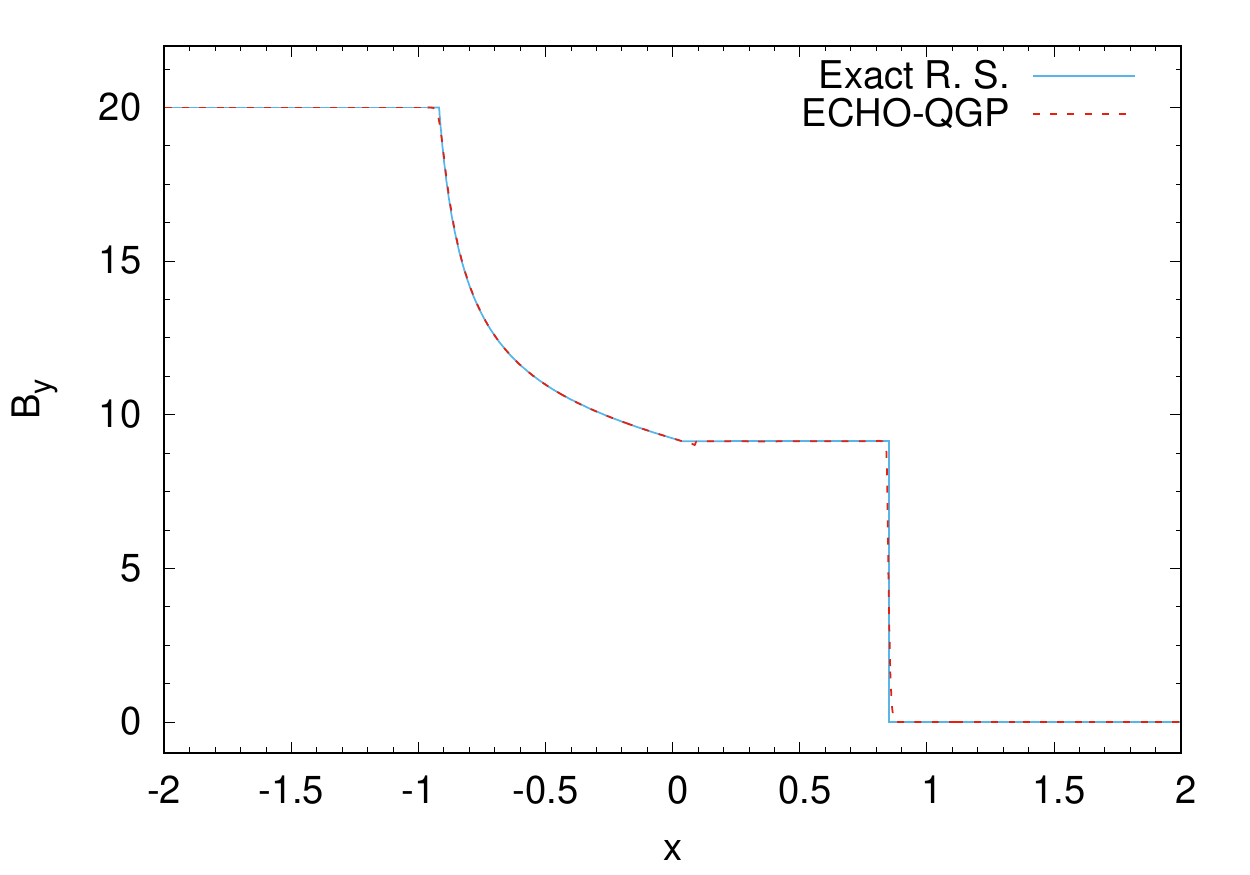}
		\includegraphics[width=0.5\textwidth]{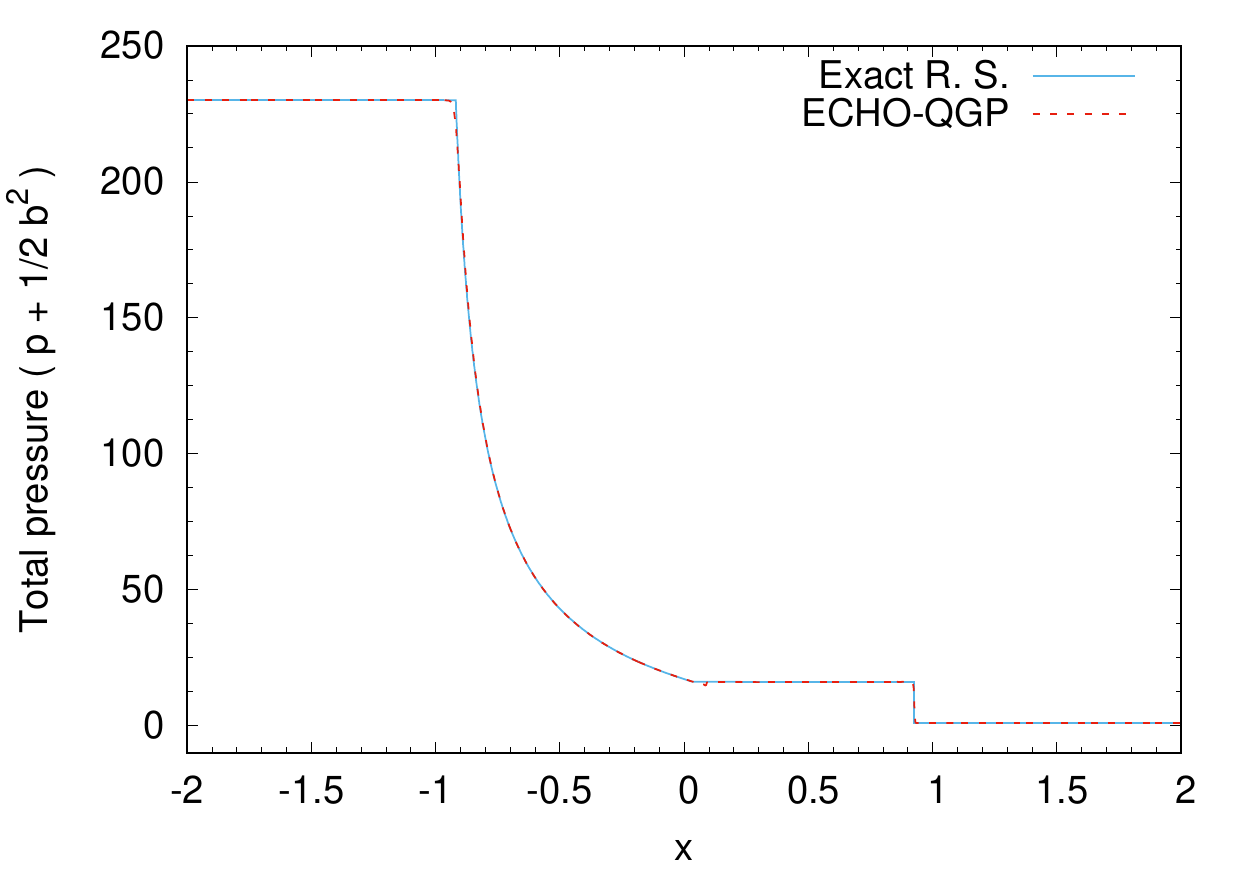}
	}
	\caption{(color online) Magnetized shock tube test for $t=4$, with the comparison of quantities computed by ECHO-QGP against the solution given by the Exact Riemann Solver by Giacomazzo and Rezzolla~\cite{giaco06}. We display the mass density $\rho$ (top left), the $v_x$ velocity component (top right), the $B_y$ magnetic-field component (bottom left) and the total pressure $p+\frac{1}{2}b^2$ (bottom right), where $b$ is the magnetic field in the comoving fluid frame.}
	\label{kom}
\end{figure*}

In order to test the shock-capturing properties of ECHO-QGP for relativistic MHD, we run a 1D shock-tube test in Minkowski coordinates comparing the numerical results against the solutions of the same problem computed by the exact Riemann solver developed by Giacomazzo and Rezzolla \cite{giaco06}. Since the cited solver works for an ideal-gas EoS, for the present test we impose
\beq
p=(\Gamma-1)(e-\rho),
\label{idgaseos}
\eeq
with an adiabatic index $\Gamma=4/3$, where $\rho = n m$ stands for the mass density in the comoving frame ($m$ is the rest mass and $n$ is the number density of the conserved species), in a situation in which particle creation/annihilation is negligible, so that $(e-\rho)$ is the thermal energy density. To employ Eq.(\ref{idgaseos}) in this test, when retrieving the primitive variables we used the same method described in Ref. \cite{ldz07}.

\begin{table}[!ht]
	\center
	\begin{tabularx}{0.45\textwidth}{l r |c| l r}
		\textbf{Left side} ($x < 0$)&$\,$&\hspace{8cm}&\textbf{Right side} ($x > 0$)& \\
		\hline
		$\rho$&1&$\,$&$\rho$&0.1\\
		$p$&30&$\,$&$p$&1\\
		$B_y$&20&$\,$&$B_y$&0\\
	\end{tabularx}
	\caption{Initial conditions for the magnetized shock tube test.}
	\label{table:ic_mag_shock_tube}
\end{table}

The initial conditions for the non-vanishing quantities are provided in \cite{kom99} and listed in Table (\ref{table:ic_mag_shock_tube}) using proper dimensionless units.

The test runs from an initial time $t\!=\!0$ to a final time $t\!=\!4$, with a grid resolution of 0.0025 (400 cells per unit of length). Results are displayed in Fig. (\ref{kom}). The comparison shows excellent agreement between the RMHD implementation in ECHO-QGP and the exact result.


\subsection{Large-amplitude CP Alfv\'{e}n-wave}

\begin{figure*}
	\hspace{-2mm}
	\includegraphics[width=0.5\textwidth]{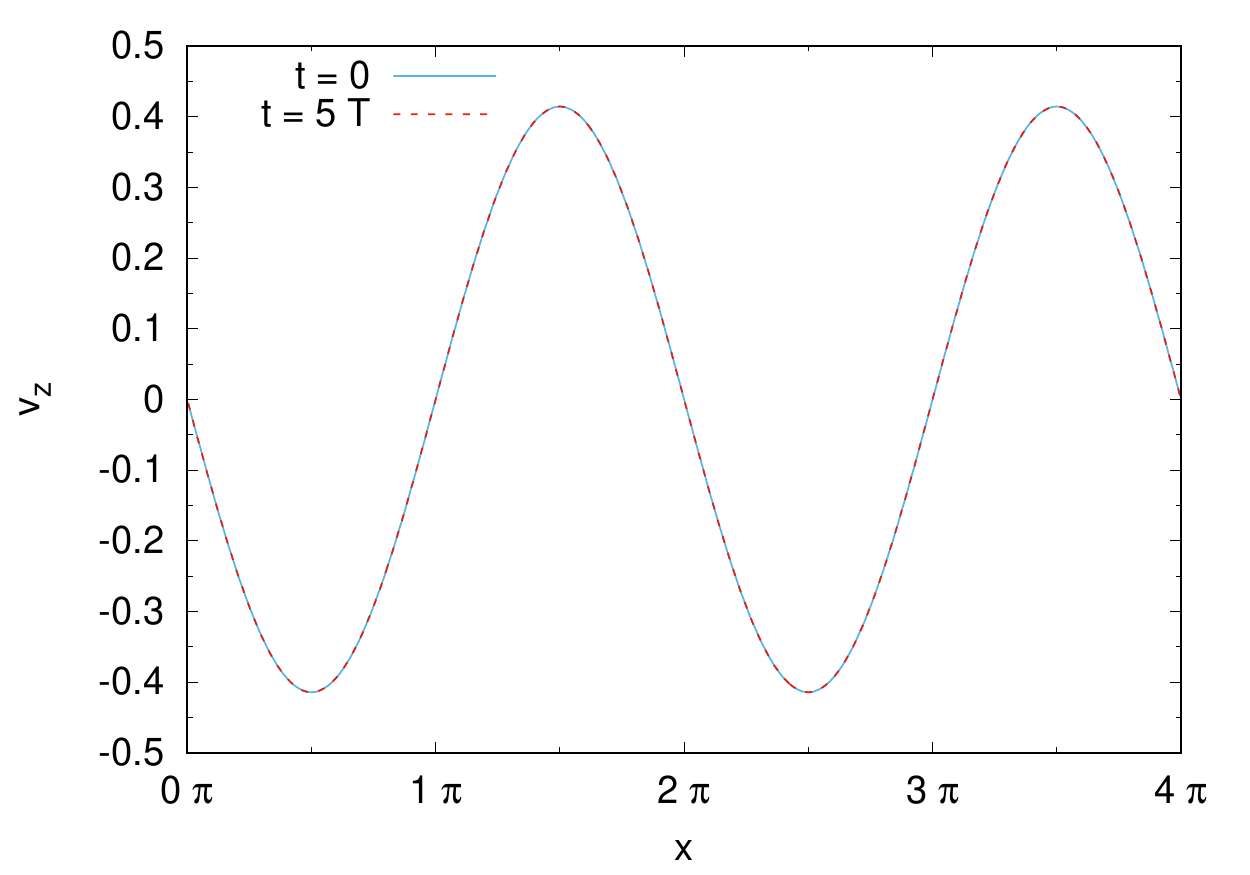}
	\includegraphics[width=0.5\textwidth]{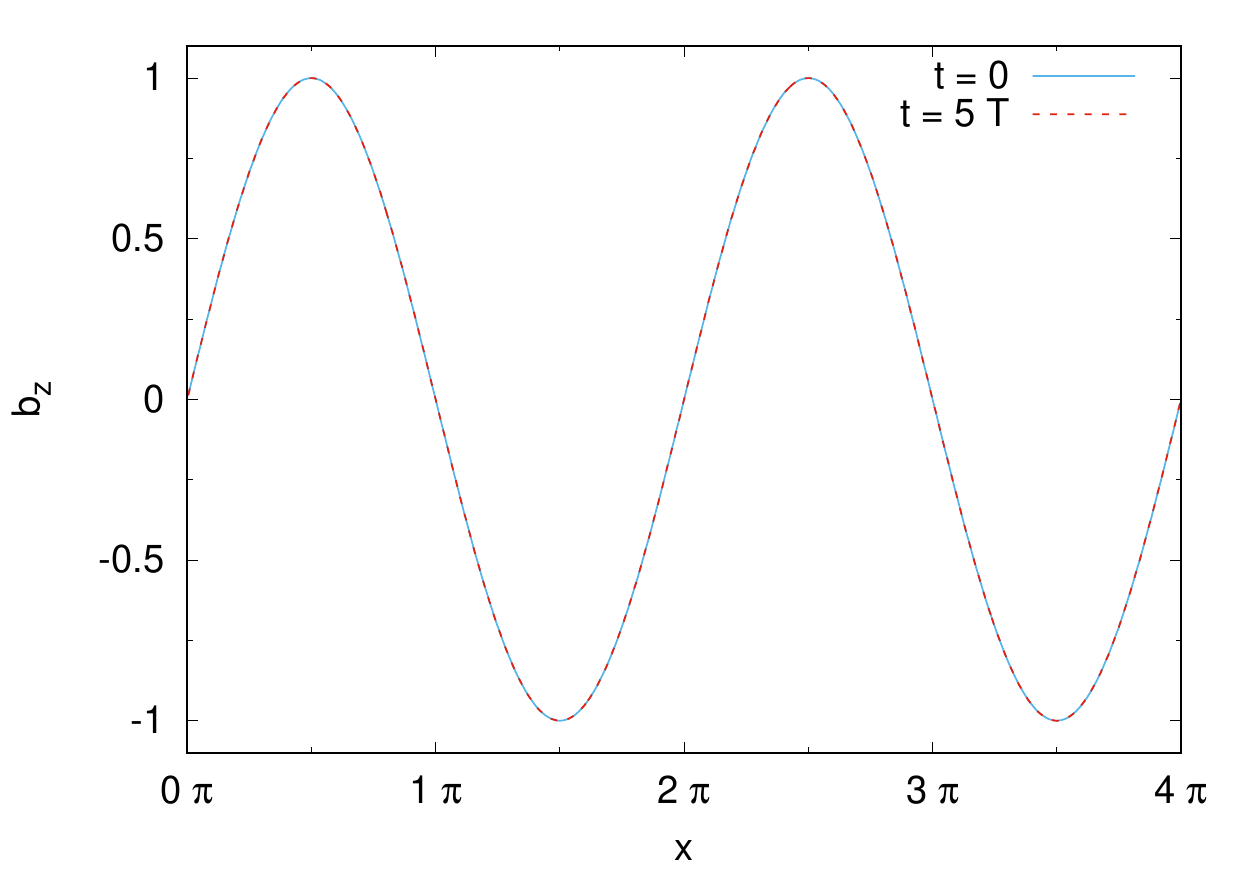}
	\caption{(color online) Circularly-polarized Alfv\'{e}n-wave test: comparison between the velocity $v_z(y=x)$ (left) and the magnetic field $B_z(y=x)$ (right) at $t=0$ and after 5 periods of the wave.}
	\label{cpwave}
\end{figure*}

A multi-dimensional relativistic MHD test with an exact\footnote{Exact in the sense, that it does not rely on the linearization of small perturbations.} solution is provided by the propagation along the diagonal of a square numerical domain of a large-amplitude Circularly Polarized (CP) Alfv\'{e}n-wave \cite{ldz07}.

We consider a Cartesian $X-Y-Z$ frame, rotated along $Z\equiv z$ in the $x-y$ plane in such a way that $X$ coincides with the diagonal $y=x$ of the numerical domain. A relativistic MHD CP Alfv\'en wave is defined by the magnetic field and velocity components
\beq
B_X = B_0, \quad B_Y=\eta B_0\cos\phi,\quad B_Z =\eta B_0\sin\phi, \nonumber
\eeq 
\beq
v_X=0, \quad v_Y=-v_A B_Y/B_0, \quad v_Z=-v_A B_Z/B_0,
\label{alfven}
\eeq
where $B_0$ is the uniform background field, the dimensionless parameter $\eta=\sqrt{B_X^2 + B_Y^2}/B_0$ sets the scale of the perturbation, and $\phi$ is the phase. For propagation along $X$ we have $\phi\!=\!k(X - v_At)$, where $k=2\pi/\lambda$ is the wave-number and the relativistic Alfv\'en velocity for arbitrary large amplitudes $\eta$ is given by \cite{ldz07}:
\beq
v_A^2 \! = \! \dfrac{2B_0^2}{ e \! + \! p + (1 \! + \! \eta^2)B_0^2 \! + \! \sqrt{[e \! + \! p + (1 \! + \! \eta^2)B_0^2]^2 - 4\eta^2B_0^4}}.
\label{alfvenspeed}
\eeq

We remind that in our ideal MHD approach the electric field is given by Eq. (\ref{eq:ohm2}) and we notice that the quantities $v^2\equiv|\vec{v}|^2\!=\!\eta^2 v_A^2$, $B^2\!=\!B_0^2(1+\eta^2)$ and $E^2\!=\!\eta^2 v_A^2 B_0^2$ are constant. Here we use the ultrarelativistic EoS $p\!=\!e/3$, where $p$ and $e$ remain constant to their initial uniform values $p_0$ and $e_0$. Notice that, as expected, for small amplitudes the Alfv\'en speed in Eq.~(\ref{alfvenspeed}) correctly reduces to the expression derived in Appendix~\ref{app:Alfven} for the linearized case. 
With the above assumptions the CP Alfv\'{e}n wave has a period $T=\lambda/v_A$, so that at time $t=nT$, with $n$ any integer number, the numerical solution is expected to assume the same configuration as at $t=0$. In the following we will consider the case of a perturbation with wavelength $\lambda=L/2$, where $L$ is the length of the diagonal of the $x-y$ domain.
We perform the test in a square numerical domain $[0,2\pi\sqrt{2}] \times [0,2\pi\sqrt{2}]$, so that $L=4\pi$, discretized with a grid of 512 x 512 cells, choosing  $p_0\!=\!e_0/3=B_0^2=1$ and also a large amplitude of the wave $\eta=1$ and a unit wave number $k=1$ (so that $\lambda=2\pi=L/2$). 

In Fig. (\ref{cpwave}) we compare the $z$ components of the velocity and of the magnetic field for $t=5T$, that is after $n=5$ periods, along the diagonal of the grid $y=x$. Neither deformations nor phase lags are observed for the depicted components as well for the other quantities not shown here. The accuracy obviously depends on an adequate numerical resolution and on the order of time and spatial integration. Further details can be found in Ref.~\cite{ldz07}. Finally, note that large-amplitude Alfv\'en waves, even if exact solutions of MHD equations, may be unstable on long timescales due to coupling with compressive modes~\cite{DelZanna01, DelZanna15}.

\subsection{Rotor test}\label{sec:Rotor}

We now describe a modified version of the 2D ``rotor'' test~\cite{ldz03,Mignone20102117}, here both in Minkowski and in Milne coordinates, using the the ultrarelativistic EoS $p\!=\!e/3$.

\begin{figure*}[!htb]
	\center
	\includegraphics[width=0.9\textwidth]{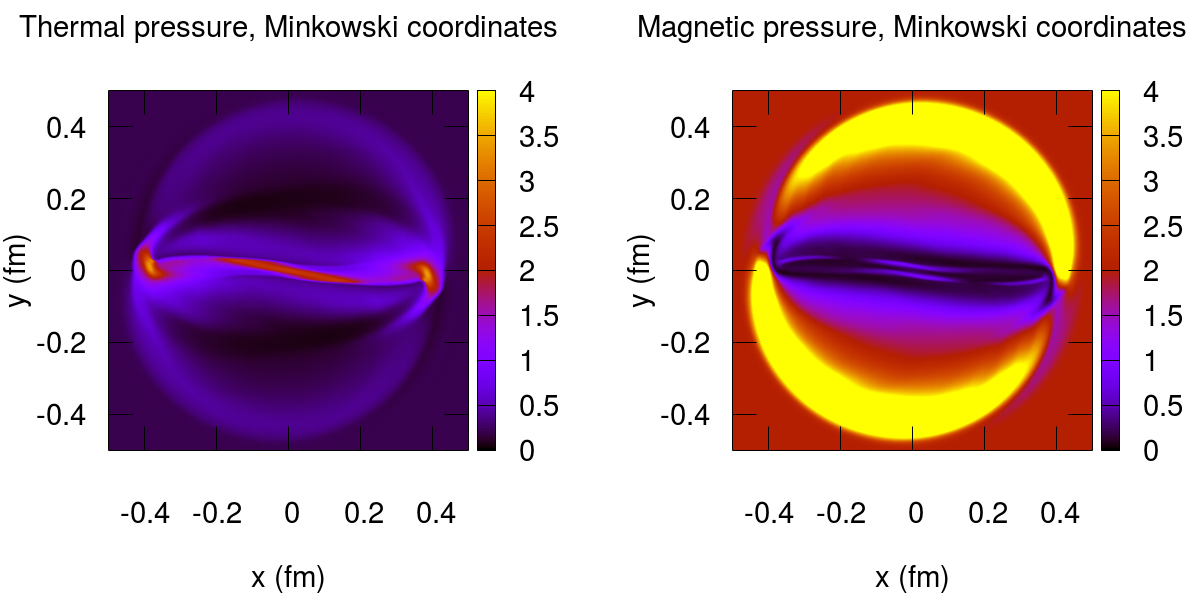}
	\caption{(color online) Results of the Rotor test in Minkowski coordinates at $t_f\!=\!1.4$ (start time was $t_i\!=\!1$), using a grid of 400x400 cells. The left plot shows the thermal pressure, the right plot shows the magnetic pressure ($(B^x B_x + B^y B_y)/2$).}
	\label{rotor_plot_mink}
\end{figure*}

An initially rigidly rotating disk of radius $r_0$ is threaded by a constant magnetic field, causing a rapid slow down of the motion. In the previous examples found in the literature the disk is denser than the surrounding medium, but, since in our case the density does not have any influence on the evolution of the system, because the EOS does not depend on it, we assume that the region inside the disk has an initial thermal pressure larger than the region outside. After this modification, the new test proposed here becomes a sort of mixture between the ``rotor'' and the ``magnetized cylindrical blast wave'' tests~\cite{ldz03}.

\begin{figure*}[!htb]
	\center
	\includegraphics[width=0.9\textwidth]{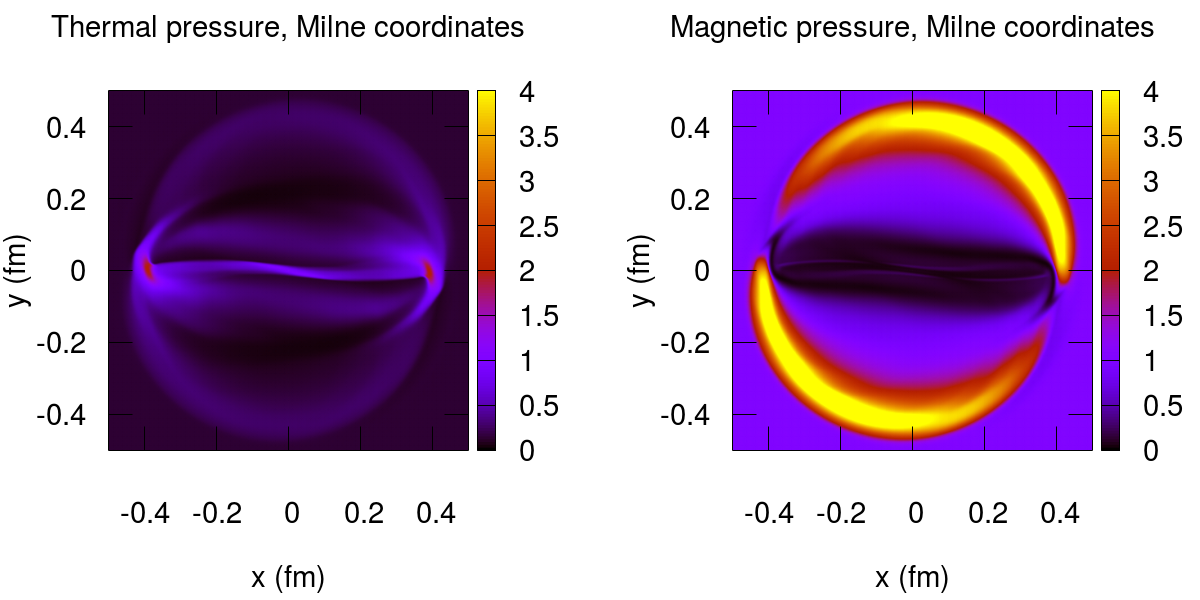}
	\caption{(color online) Results of the Rotor test in Milne cordinates at $t_f\!=\!1.4$ (start time was $t_i\!=\!1$), using a grid of 400x400 cells. The left plot shows the thermal pressure, the right plot shows the magnetic pressure ($(B^x B_x + B^y B_y)/2$).  We remind that in Milne coordinates $v^{\eta}\!=\!0$ $\Leftrightarrow$ $v^z=z/t$, implying that this case describes the evolution of a system which is different from the other one in Minkowski coordinates.}
	\label{rotor_plot_milne}
\end{figure*}

The initial velocity of the fluid is null outside of the disk, while inside the disk its components are:
\[ 
\begin{dcases}
v^x\,=\,\dfrac{\omega\,y}{r_0}\\
v^y\,=\,-\dfrac{\omega\,x}{r_0}\\
v^z\,=\,0
\end{dcases} \]
in the case of Minkowski coordinates, while, in Milne coordinates, $v^z$ is substituted by $v^{\eta}\!=\!0$, which amounts to assume a longitudinal Bjorken expansion $v^z=z/t$.

\begin{table}
	\center
	\begin{tabularx}{0.45\textwidth}{l l l}
		Parameter&Description&Value\\
		\hline
		$r_0$&disk radius&0.1\\
		$\omega$&Rot. speed param.&0.995\\
		$B^x$&(everywhere)&2\\
		$B^y$&(everywhere)&0\\
		$B^z$&(everywhere)&0\\
		$p$&thermal pressure ($r\,\le \,r_0$)&5\\
		$p$&thermal pressure ($r\,>\,r_0$)&1\\
		$t_i$&start time&1\\
		$t_f$&end time&1.4\\
	\end{tabularx}
	\caption{Values of the parameters used in the rotor test.}
	\label{table:rotor_params}
\end{table}

The values of the parameters chosen for the test are listed in Table (\ref{table:rotor_params}).

The major difference between the results in the two coordinate systems is the decay of the thermal and magnetic pressures in the case of Milne coordinates, which occurs in every region of the grid, due the longitudinal expansion of the system. Then, in both cases we observe a compression wave, due to the larger initial inner pressure and due to the motion of rotation of the disk, forged into an asymmetric shape by the effects of the magnetic field.

\subsection{Bjorken flow}\label{sec:Bjorken}
This test consists in a comparison with the analytical solution for the temporal evolution of a one-dimensional boost-invariant flow, obtained extending the model by J.D. Bjorken~\cite{bj83} to the case of transverse MHD~\cite{roy15}.\\ 
We consider the relativistic flow along the $z$-direction of an ideal magnetized fluid, with pressure $p$ and energy density $e$, related by the ultrarelativistic EoS $p\!=\!e/3$, both constant in the transverse $x\!-\!y$ plane and independent from the space-time rapidity $\eta_s$ (one employs Milne coordinates).
For the flow profile one considers a longitudinal boost-invariant Hubble-law expansion $v^z\!=\!z/t$, leading to a four velocity $u^\mu\!=\!(\cosh\eta_s,0,0,\sinh\eta_s)$. In Milne coordinates the fluid velocity reads simply $u^\mu\!=\![1,0,0,0]$, so that for the comoving derivative and the expansion rate one has $D\!=\!\partial_\tau$ and $\theta\!=\!1/\tau$.
The transverse MHD hypothesis, i.e. the assumption of having a magnetic field $b_\mu\!=\!(0,b_x,b_y,0)$ orthogonal to the fluid velocity $u^\mu$, so that $u^\mu b_\mu=0$, allows one to derive from Eq.~(\ref{eq:energy3}) the energy-conservation equation~\cite{roy15}
\begin{equation}
\partial_{\tau} \left(e+\frac{{b}^{2}}{2}\right)+\frac{e+p+{b}^{2}}{\tau}=0.
\label{bj_enlaw}
\end{equation}

However, under the hypothesis of infinite conductivity, one has also from Eq.(\ref{eq:energy2})
\beq
\partial_{\tau}\,e+\frac{e+p}{\tau}=0.
\eeq
This allows one to obtain the evolution equation for the magnetic field:
\beq
\partial_\tau b+\frac{b}{\tau}=0.
\eeq

\begin{figure}
	\center
	\includegraphics[width=0.48\textwidth]{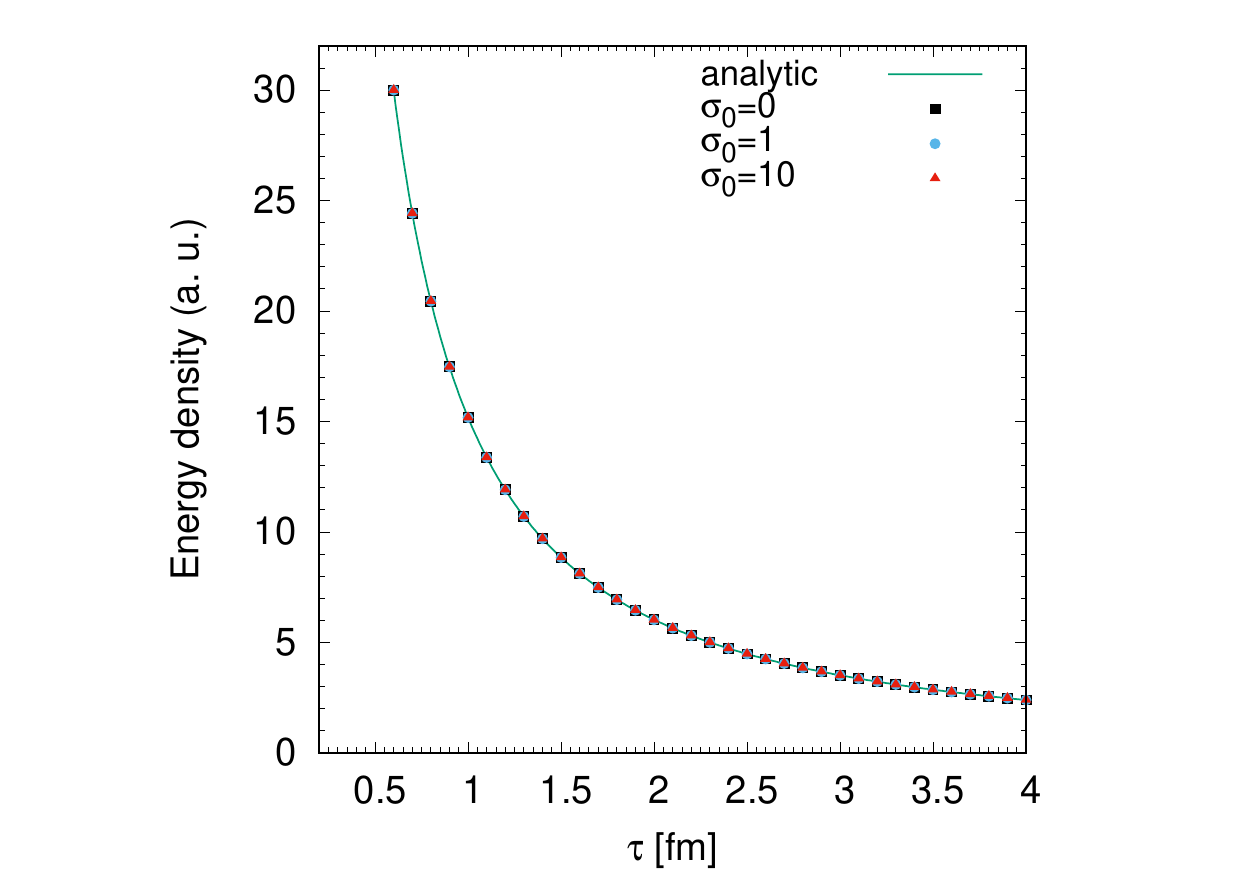}
	\caption{(color online) Evolution of the energy density in arbitrary units for different values of the initial magnetization $\sigma_0$ in comparison to the analytical result .}
	\label{bj_endens}
\end{figure}

\begin{figure}
	\center
	\includegraphics[width=0.48\textwidth]{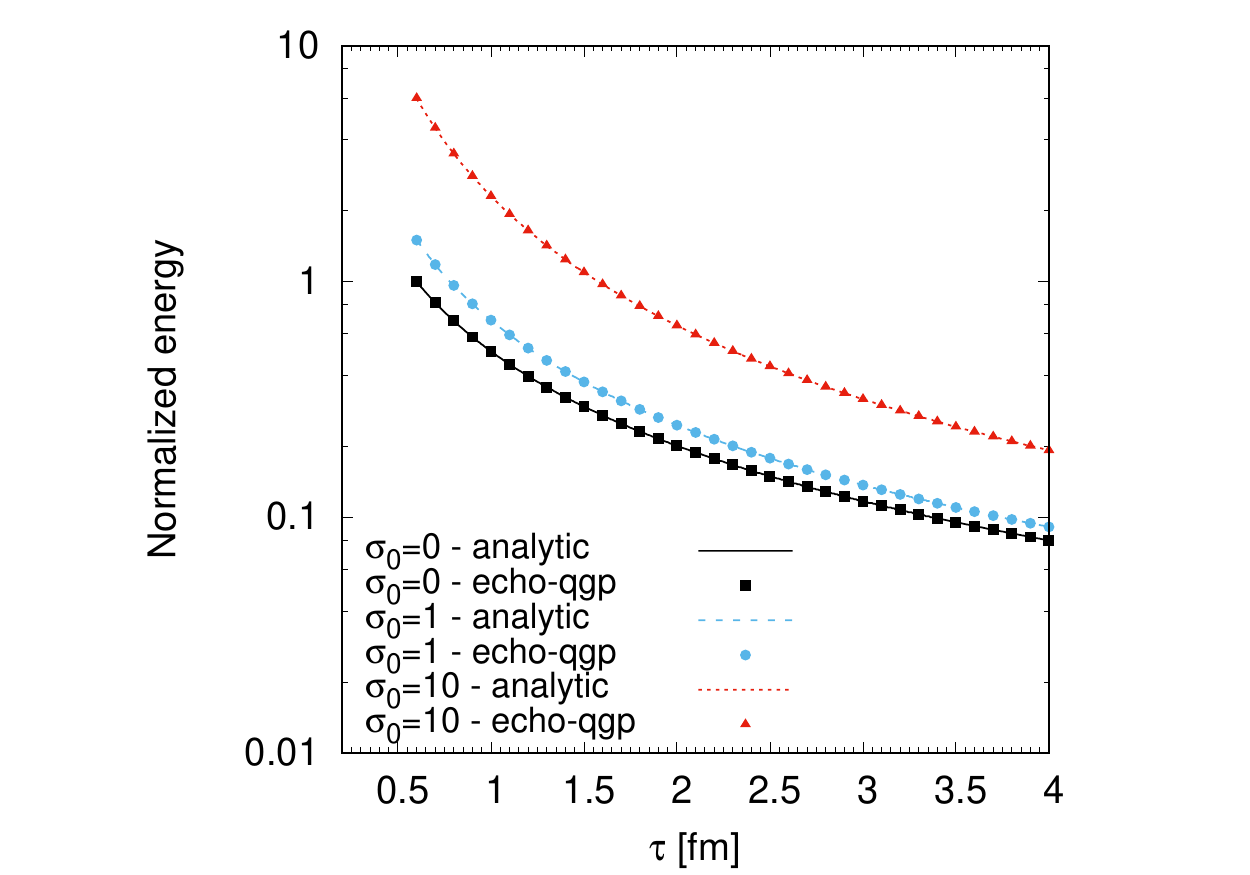}
	\caption{(color online) Evolution of the normalized total energy density $(e+b^2/2)/e_0$ in comparison to the analytical result.}
	\label{bj_norm_endens}
\end{figure}

Considering the case of an ultrarelativistic $p=e/3$ EoS, it is possible to derive from the above the time evolution of the energy density and of the magnetic field:
\begin{eqnarray}
e(\tau)&=&e_0\left(\dfrac{\tau_0}{\tau}\right)^{4/3}
\end{eqnarray}
and
\begin{eqnarray}
b(\tau)&=&b_0 \dfrac{\tau_0}{\tau} .
\end{eqnarray}
Notice that, in an ideal plasma, due to the flux-freezing condition, the magnetic field decreases according to the same law as the conserved charges or of the entropy.

We perform the test for three different values of the initial magnetization $\sigma_0\!=\!b_0^2/e_0$: 0, 1 and 10 (in adimensional units).
The comparison with the analytic results (shown in Figures (\ref{bj_endens}) and (\ref{bj_norm_endens})) shows perfect agreement between the simulation and the exact solution.

\subsection{Self-similar expansion in vacuum}\label{sec:Lyutikov}
With the purpose of performing a non-trivial validation of our numerical code, here we consider an exact solution of the so-called transverse RMHD equations, namely a situation in which a hot magnetized plasma flows along one direction, with the magnetic field perpendicular to the flow. Without loss of generality we can adopt a Minkowskian flat space in Cartesian coordinates and take the fluid flowing along the $z$-axis, while the magnetic field having only $x$-component
\begin{equation}
u^\mu=\gamma(1,0,0,v),\quad b^\mu=(0,b,0,0)=(0,B/\gamma,0,0).
\end{equation}
Within ideal RMHD, we recall that the relation between the magnetic field in the comoving ($b^\mu$) and laboratory ($\B$) frames is given by Eq.~(\ref{eq:bvsB_id}):
\beq
b^\mu=[\gamma(\v\!\cdot\!\B),\B/\gamma+\gamma(\v\!\cdot\!\B)\v],
\eeq 
and if all quantities are constant in the transverse plane, the set of equations reduces to
\begin{subequations}
	\begin{align}
	D\left(e+{b^2}/{2}\right)+(e+p+b^2)\theta=0, \\
	(e+p+b^2)D u^\mu+\nabla^\mu (p+b^2/2)=0.
	\end{align}
\end{subequations}
The above equations have to be solved together with the one providing the evolution of the magnetic field in the plasma
\beq
\partial_t\B = - \nab\times\E,
\eeq
which, in ideal MHD where $\E=-\v\times\B$, leads to
\beq
(\partial_t+\v\!\cdot\!\nab)\B=(\B\!\cdot\!\nab)\v-\B(\nab\!\cdot\!\v). \label{eq:Faraday_id}
\eeq
Writing explicitly the derivatives one gets:
\begin{subequations}
	\begin{align}
	(\partial_t+v\,\partial_z)\left(e+{b^2}/{2}\right)+\gamma^2(e+p+b^2)(v\,\partial_t+\partial_z)v=0\\
	(v\,\partial_t+\partial_z)(p+b^2/2)+\gamma^2(e+p+b^2)(\partial_t+v\,\partial_z)v=0
	\end{align}
\end{subequations}
and
\beq
(\partial_t+v\,\partial_z)B=-B(\partial_z v).
\eeq 

We now wish to address the case of a plasma, initially at rest, with magnetic field, pressure, energy and entropy density $b_0$, $p_0$, $e_0$ and $s_0$ for $z<0$ and vanishing on the right. We want to study how the system evolves in time, extending the study performed in~\cite{PhysRevE.85.026401} to the case of an ultra-relativistic plasma of massless particles. 
For this purpose, it is useful to introduce the self-similar variable $\xi\equiv z/t$, which allows one to rewrite the equations as:
\begin{subequations}
	\begin{align}
	(v-\xi)\frac{d}{d\xi}\left(e+{b^2}/{2}\right)+\gamma^2(e+p+b^2)(1-v\,\xi)\frac{dv}{d\xi}=0\\
	(1-v\,\xi)\frac{d}{d\xi}(p+b^2/2)+\gamma^2(e+p+b^2)(v-\xi)\frac{dv}{d\xi}=0
	\end{align}
\end{subequations}
In Ref.~\cite{PhysRevE.85.026401} the system was closed by combining the induction equation for the magnetic field with the one for mass conservation. Actually, in the case of heavy-ion collisions, such a choice would not be meaningful, since one deals with an ultra-relativistic plasma of massless particles, in which particle-antiparticle pairs are continuously created/annihilated. However, in the absence of dissipative effects, one can replace the conservation equation for the mass with the one for the entropy. One can write the conservation law $d_\mu s^\mu = \partial_\mu s^\mu=0$ for the entropy current
\beq
s^\mu\equiv s u^\mu=s\gamma\,(1,\v)\equiv \tilde{s}\,(1,\v).
\eeq
Entropy conservation can be expressed by Eq.~(\ref{eq:entro}) or, here more conveniently, in terms of its density in the laboratory frame:
\beq
(\partial_t+\v\!\cdot\!\nab)\tilde{s}=-\tilde{s}\,\nab\!\cdot\!\v\label{eq:entro_lab}
\eeq
Introducing the Lagrangian derivative $d/dt\equiv(\partial_t+\v\!\cdot\!\nab)$ and combining Eqs.~(\ref{eq:Faraday_id}) and (\ref{eq:entro_lab}) one gets:
\beq
\frac{d}{dt}\left(\frac{\B}{\tilde{s}}\right)=\frac{1}{\tilde{s}}(\B\!\cdot\!\nab)\,\v.
\eeq
In the transverse one-dimensional MHD case we are addressing one has then:
\beq
{\frac{d}{dt}\left(\frac{B}{\tilde{s}}\right)=\frac{d}{dt}\left(\frac{b}{s}\right)=0}\quad\longrightarrow\quad
b(db)=b^2\frac{ds}{s}\label{eq:b/s}
\eeq
This allows one to rewrite the set of RMHD equations as (the prime index denotes the derivative with respect to the self-similar variable $\xi$)
\begin{subequations}
	\begin{align}
	(v-\xi)\left(e'+b^2\frac{s'}{s}\right)+\gamma^2(e+p+b^2)(1-v\,\xi)v'=0\\
	(1-v\,\xi)\left(p'+b^2\frac{s'}{s}\right)+\gamma^2(e+p+b^2)(v-\xi)v'=0\\
	s'=-\frac{1-v\,\xi}{v-\xi}\gamma^2 s v'\,,\label{eq:s'}
	\end{align}
\end{subequations}
The equation for the entropy, together with the rather general EoS $p=c_s^2 e$ (here $c_s$ is the sound speed), leads to
\begin{subequations}
	\begin{align}
	&(v-\xi)e'+\gamma^2(e+p)(1-v\,\xi)v'=0\\
	&(v-\xi)(1-v\,\xi)c_s^2 e'\nonumber\\
	&+\gamma^2
	\left[(e+p+b^2)(v-\xi)^2-(1-v\,\xi)^2 b^2\right]v'=0
	\end{align}
\end{subequations}
The system has a non-trivial solution only if the determinant vanishes, i.e. if
\beq
(1-v\,\xi)^2 c_s^2(e+p)=(e+p)(v-\xi)^2-(1-v^2)(1-\xi^2)b^2.
\eeq
A rarefaction wave propagates from the outside inside the plasma. The position of the rarefaction front, characterized by a vanishing value of the fluid velocity $v=0$ and with all the other quantities equal to their initial unperturbed values is given by
\beq
c_s^2(e_0+p_0)=(e_0+p_0)\xi_{\rm rw}^2-(1-\xi^2_{\rm rw})b^2.
\eeq
One gets then
\beq
\xi^2_{\rm rw}=\frac{(e_0+p_0)c_s^2+b_0^2}{e_0+p_0+b_0^2},
\eeq
which, in the case of and ideal ultrarelativistic gas EoS, reduces to
\beq
{\xi^2_{\rm rw}=\frac{(4/3)p_0+b_0^2}{4p_0+b_0^2}}\,,
\eeq
in agreement with what obtained for the \emph{fast magnetosonic speed} in Eq.~(\ref{eq:fast1}) of \ref{app:fast}. Hence, with the initial condition we chose, the position of the rarefaction front propagates backwards with a velocity equal to the fast magnetosonic speed: $z_{\rm rf}(t)=-c_f\,t$.

We now look for an explicit solution written in terms of the ratio ${\cal B}$ between the initial thermal and magnetic pressure. We will try to follow an approach as close as possible to the one employed by Lyutikov and Hadden~\cite{PhysRevE.85.026401}.
In the case of an ideal ultra-relativistic plasma one has $p\sim T^4$ and $s\sim T^3,$ so that
\beq
p=p_0\left(\frac{s}{s_0}\right)^{4/3}\quad\longrightarrow\quad p'=\frac{4}{3}p\left(\frac{s'}{s}\right)\label{eq:Pvss}
\eeq
One gets then
\beq
(1-v\,\xi)\left(\frac{4}{3}p+b^2\right)\frac{s'}{s}+\gamma^2(4p+b^2)(v-\xi)v'=0.
\eeq
Exploiting Eq.~(\ref{eq:s'}) one obtains
\beq
(4p+b^2)(v-\xi)^2-\left(\frac{4}{3}p+b^2\right)(1-v\,\xi)=0.
\eeq
In the approach by Lyutikov (generalized to our ultra-relativistic case) one writes the above equation in terms of the parameter and variable
\beq
{{\cal B}\equiv\frac{p_0}{b_0^2/2}}
\quad{\rm and}\quad {s_1\equiv\frac{s}{s_0}}
\eeq
One has then, from Eqs.~(\ref{eq:b/s}) and (\ref{eq:Pvss}) 
\beq
p=p_0\left(\frac{s}{s_0}\right)^{4/3}\!\!\!={\cal B}\,\frac{b_0^2}{2}\,s_1^{4/3}
\eeq
and
\beq
b^2=b_0^2\,\frac{b^2}{b_0^2}=b_0^2\,\frac{s^2}{s_0^2}=b_0^2 s_1^2.
\eeq

Hence, we get
\beq
(2{\cal B}+s_1^{2/3})(v-\xi)^2-\left(\frac{2}{3}{\cal B}+s_1^{2/3}\right)(1-v\,\xi)=0,
\eeq
which we can recast as
\beq
\begin{aligned}
	&s_1^{2/3}(1-v^2)(1-\xi^2)+\frac{2}{3}{\cal B}[1+4v\,\xi-3\xi^2+v^2(\xi^2-3)]\\
	&=0
\end{aligned}
\eeq
The latter is equivalent to Eq. (6) in the paper by Lyutikov, except that now it depends only on the parameter ${\cal B}$ (thermal pressure and particle/entropy density are not independent variables in an ultra-relativistic plasma) and it is does not include the term arising from the mass density.

The above equations can be equivalently written in terms of the variables
\beq
\delta_v\equiv\sqrt{\frac{1+v}{1-v}},\quad \delta_\xi\equiv\sqrt{\frac{1+\xi}{1-\xi}}.
\eeq
One obtains
\begin{subequations}
	\begin{align}
	\delta_v^2\delta_\xi^2 s_1^{2/3}-\frac{1}{3}{\cal B}[\delta_v^4-4\delta_v^2\delta_\xi^2+\delta_\xi^4]=0\\
	(\delta_v^2+\delta_\xi^2)s_1\frac{\partial \delta_v}{\partial \delta_\xi}+
	\delta_v(\delta_v^2-\delta_\xi^2)\frac{\partial s_1}{\partial \delta_\xi}=0
	\end{align}
\end{subequations}
From the first equation we define
\beq
\frac{\delta^2_v}{\delta^2_\xi}\equiv f^2(s_1)\equiv\frac{(4{\cal B}+3s_1^{2/3})\pm \sqrt{(4{\cal B}+3s_1^{2/3})^2-4{\cal B}^2 }}{2{\cal B}}
\eeq
From the second equation one gets then
\beq
\frac{\partial \ln\delta_\xi}{\partial s_1}=\frac{f(s_1)(1-f^2(s_1))-s_1 f'(s_1)(1+f^2(s_1))}{s_1\,f(s_1)(1+f^2(s_1))}
\eeq
The latter can be easily integrated, obtaining
\beq
\ln\frac{\delta_\xi(s_1)}{\delta_{\xi_0}}=
\int_1^{s_1}\!\!\!d\alpha \frac{f(\alpha)(1-f^2(\alpha))-\alpha f'(\alpha)(1+f^2(\alpha))}{\alpha\,f(\alpha)(1+f^2(\alpha)},
\eeq
where $\delta_{\xi_0}$ can be fixed through the initial condition, namely the development of a left-propagating rarefaction-wave, with velocity equal to the fast magnetosonic speed:
\beq
\delta_{\xi_0}=\sqrt{\frac{1-c_{f,0}}{1+c_{f,0}}},\quad{\rm where}\quad
c^2_{f,0}=\frac{2{\cal B}+3}{3(2{\cal B}+1)}.
\eeq

\begin{figure}
	\center
	\includegraphics[width=0.45\textwidth]{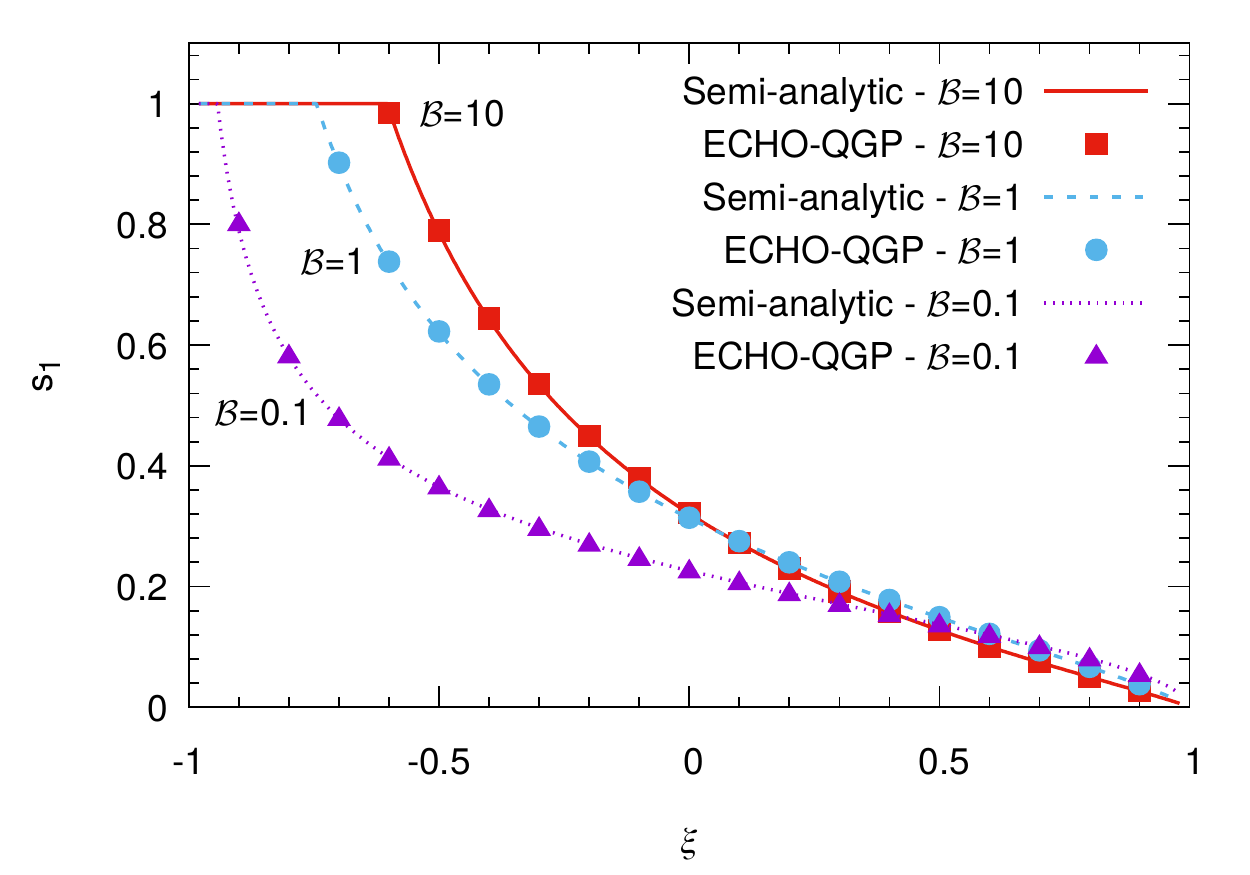}
	\caption{(color online) Self-similar expansion into vacuum test, comparison of the ECHO-QGP results with the semi-analytic solution computed with Mathematica~\cite{mathematica9}. The graph shows $s_1=s/s_0$ vs $\xi=z/t$ at $t=20$ for three different values ( 10, 1 and 0.1 from top to bottom) of the $\cal B=2 p_0/B_0^2$ parameter. We used a grid of 801 cells, the reconstruction algorithm MPE5, the approximate Riemann solver HLL and the time integration algorithm was a second order Runge-Kutta. The initial pressure was: left side ($z \leq 0$) $p_0 = 1000$, right side ($z > 0$) $p_0 = 5\cdot 10^{-5}\approx 0$ (due to numerical reasons).}
	\label{plot:selfsim}
\end{figure}

In Fig. (\ref{plot:selfsim}) we display a comparison between the above semi-analytic solution and the numerical result provided by our code. The graph shows $s_1=s/s_0$ vs $\xi=z/t$ at $t=20$ for three different values ( 10, 1 and 0.1 ) of the $\cal B=2 p_0/B_0^2$ parameter. We used a grid of 801 cells, reconstruction algorithm: MPE5, approximate Riemann solver: HLL, time integration algorithm: second order Runge-Kutta. Initial pressure was: left side ($z \leq 0$) $p_0 = 1000$, right side ($z > 0$) $p_0 = 5\cdot 10^{-5}\approx 0$ (due to numerical reasons, since ECHO-QGP cannot run with true null pressure). Again we observe excellent agreement between the numerical implementation and the analytical results for a large variety of parameters.

\section{Results of RMHD simulations for HIC}\label{sec:results}
We plan to present a more extensive study of the QGP evolution in a subsequent article, nevertheless here we present some preliminary results to evaluate the impact that the interplay between magnetic field and hydro evolution may have on some experimental observables. Although the whole 3D+1 formalism has been already implemented into the code, for simplicity here we will show a basic 2D+1 application to Heavy Ion Collisions.
\subsection{Setup}
We consider Au-Au collisions at $\sqrt{s_{NN}}$=200 GeV. The initial conditions are modelled with the optical Glauber model~\cite{glauber59,DelZanna:2013eua}.
In this framework, the initial energy density distribution $e$ in the transverse plane is given by:
\be
e(\tau_0,\x;b)=e_0\left[(1-\alpha_H)\frac{n_{\rm part}(\x;b)}{n_{\rm
		part}(\0;0)}+\alpha_H\frac{n_{\rm coll}(\x;b)}{n_{\rm coll}(\0;0)}\right], 
\label{endens}
\ee
where $e_0$ is the value of $e$ at $\x\!=\!\0$ and $b\!=\!0$, $\x$ the coordinates in the transverse plane and $b$ the impact parameter.
One defines the nuclear thickness function as: 
\beq
\widehat{T}_{A/B}(\x)\equiv\int_{-\infty}^{\infty}\!\!\!dz\, \rho_{A/B}(\x,z),
\label{thick_func}
\eeq
with
\beq
\int_{-\infty}^{\infty}\widehat{T}_{A/B}(\x)\,d\x\!=\!1
\label{thick_func_norm_condition}
\eeq
where $\rho_{A/B}(\x,z)$ is the Wood-Saxon nuclear density distribution for the nuclei A and B. One obtains then the density of participants $n_{\rm p}(\x;b)\equiv n_{\rm p}^A(\x;b)+n_{\rm p}^B(\x;b)$ from:
\beqa
n_{\rm p}^A(\x;b)& \! = \! A\,\widehat{T}_{A}(\x+\b/2)
\left\{1 \! - \! [1\! -\! \widehat{T}_{B}(\x-\b/2)\sigma^{\rm in}_{NN}]^B\right\} ,
\nonumber\\ 
n_{\rm p}^B(\x;b)& \! = \!B\,\widehat{T}_{B}(\x-\b/2) 
\left\{1\! - \! [1\!-\! \widehat{T}_{A}(\x+\b/2)\sigma^{\rm in}_{NN}]^A\right\},
\nonumber\\ 
\eeqa
and the number density of binary collisions in the transverse plane as:
\beq
n_{\rm c}(\x;b)=AB\,\sigma^{\rm in}_{NN}\,\widehat{T}_{A}(\x+\b/2)\widehat{T}_{B}(\x-\b/2),
\eeq
where $\sigma^{\rm in}_{NN}$ is the inelastic nucleon-nucleon cross-section.
Since ECHO-QGP is not able to run with null energy density or if the thermal pressure is much smaller than the magnetic pressure, to ensure the stability of the code, we increase the initial energy density distribution by an additional small amount $e_{min}$, negligible from the point of view of the dynamics of the system.
We adopt Milne coordinates and we assume boost invariance along the $\eta$ direction. 
The velocity components of the fluid are all null at the initial time $\tau_0$, i.e. $v^x\!=\!v^y\!=v^\eta\!=\!0$.\\ 
We compute the initial magnetic field following the approach adopted by K. Tuchin~\cite{Tuchin:2013apa}, i.e. we consider a magnetic field produced by an electric charge $e$ moving parallel to the $z$-axis with a speed $v$ having a Lorentz factor $ \gamma \gg 1$ as measured in the laboratory frame by an observer located at $\r= z\unitvec z + \b$, where $\b$ is the distance from the $z$-axis in the transverse plane ( $\b \cdot \unitvec z = 0 $). We also assume a constant permittivity $\epsilon=1$, a constant permeability $\mu=1$, a constant finite electrical conductivity $\sigma$. Under these assumptions, it can be shown that the magnetic field $\vec{B}=B(t,\r)\unitvec \phi$ is given by:
\begin{multline}
B(t,\r)=\dfrac{e (\hslash c)^{\frac{3}{2}}}{2\pi\sigma} 
\bigintss_0^\infty \dfrac{J_1(k_\bot b)k_\bot^2}{\sqrt{1+\frac{4k_\bot ^2 (\hslash c)^2}{\gamma^2\sigma^2}}}
\cdot\\
\exp\left\{ \dfrac{\sigma \gamma^2 x_\pm }{2 (\hslash c)} \left( 1- \sqrt{1+\dfrac{4k_\bot^2 (\hslash c)^2}{\gamma^2\sigma^2}} \right)\right\}\,dk_\bot
\label{eq:tuchin_base}
\end{multline}
where $x_\pm = t \pm v/z$ and $e\!=\!\sqrt{4\pi\alpha}$, $\alpha$ being the fine structure constant. We mention that the $\vec{B}$ field has dimensions [$\textrm{GeV}^{1/2}\textrm{fm}^{-3/2}$], so that $B^2$ has the same dimensions as the pressure, i.e. [$\textrm{GeV}/\textrm{fm}^{3}$].

Then, we approximate the electric charge distribution inside the two colliding nuclei as being uniform and spherical and we perform an integration over it to get the total magnetic field in each point of our computational grid. We assume that the motion and the distribution of the electric charges are unaffected by the collision between the nuclei. A detailed description of the whole procedure can be found in Ref.~\cite{Tuchin:2013apa}.
Since at the moment our code is not able to handle configurations where the magnetic pressure is much larger than the thermal pressure, which is the case in regions outside the fireball, where the initial energy density is less than 30 $\textrm{MeV}/\textrm{fm}^3$ we rescale the magnetic field so that the ratio between the magnetic and the thermal pressure does not exceed 0.1. This procedure does not affect the final results because at such low temperature there is no participating QCD matter and the hydrodynamic description of the medium would cease to be valid anyway.

\begin{table}
	\center
	\begin{tabularx}{0.45\textwidth}{l l l}
		Parameter&Description&Value\\
		\hline
		b & impact parameter & $10$ fm\\
		$\tau_0$  & initial time & $0.4$ fm/c\\
		$e_{f.o.}$ & freezeout energy density. & $150\,\textrm{MeV}/\textrm{fm}^3$\\
		$\epsilon_0$ & max. en. dens. & $55.\,\textrm{GeV}/\textrm{fm}^3$\\
		$\epsilon_{min}$ & min. en. dens. & $0.1.\,\textrm{MeV}/\textrm{fm}^3$\\
		$\sigma_{in}$ & inel. cross sect. & $40$ mb\\
		$\alpha_H$ & collision hardness & $0.05$\\
		EoS & equation of state & $p=e/3$\\
		\vspace*{2mm}
	\end{tabularx}
	\caption{Values of the parameters used in the setup of the 2D+1 RMHD simulations of heavy-ion collisions.}
	\label{tabinpar}
\end{table}

\begin{figure}
	\center
	\includegraphics[width=0.45\textwidth]{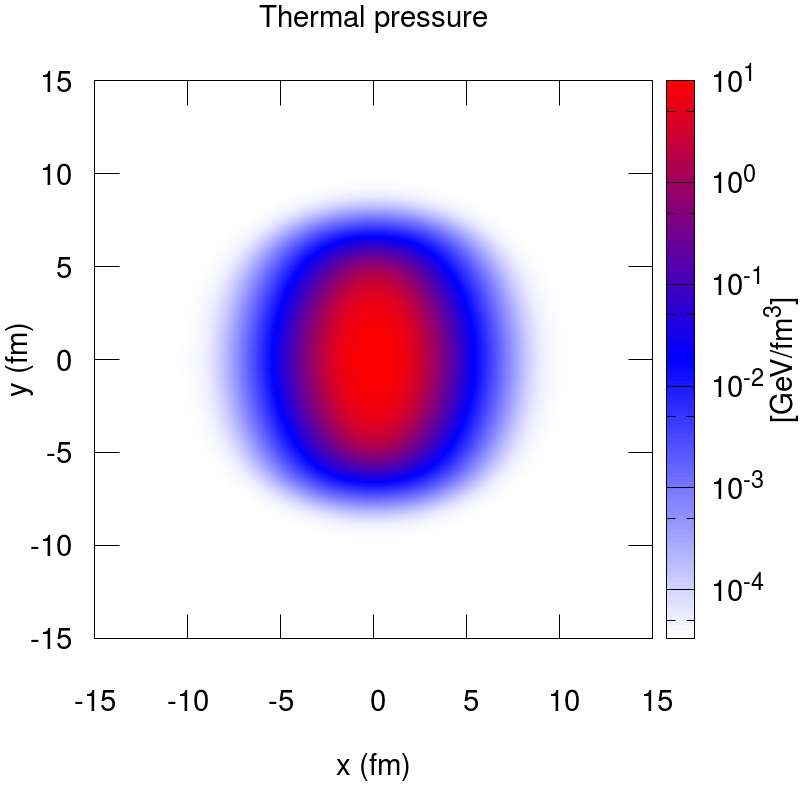}
	\caption{(color online) The initial spatial pressure distribution in the transverse plane, obtained using the geometrical Glauber model given by Eq. (\ref{endens}) with the parameters listed in Table (\ref{tabinpar}). The parameters are for the reaction Au+Au, b=10 fm at $\sqrt{s_{NN}}$=200 GeV.}
	\label{initial_pressure}
\end{figure}

\begin{figure*}
	\center
	\includegraphics[width=0.9\textwidth]{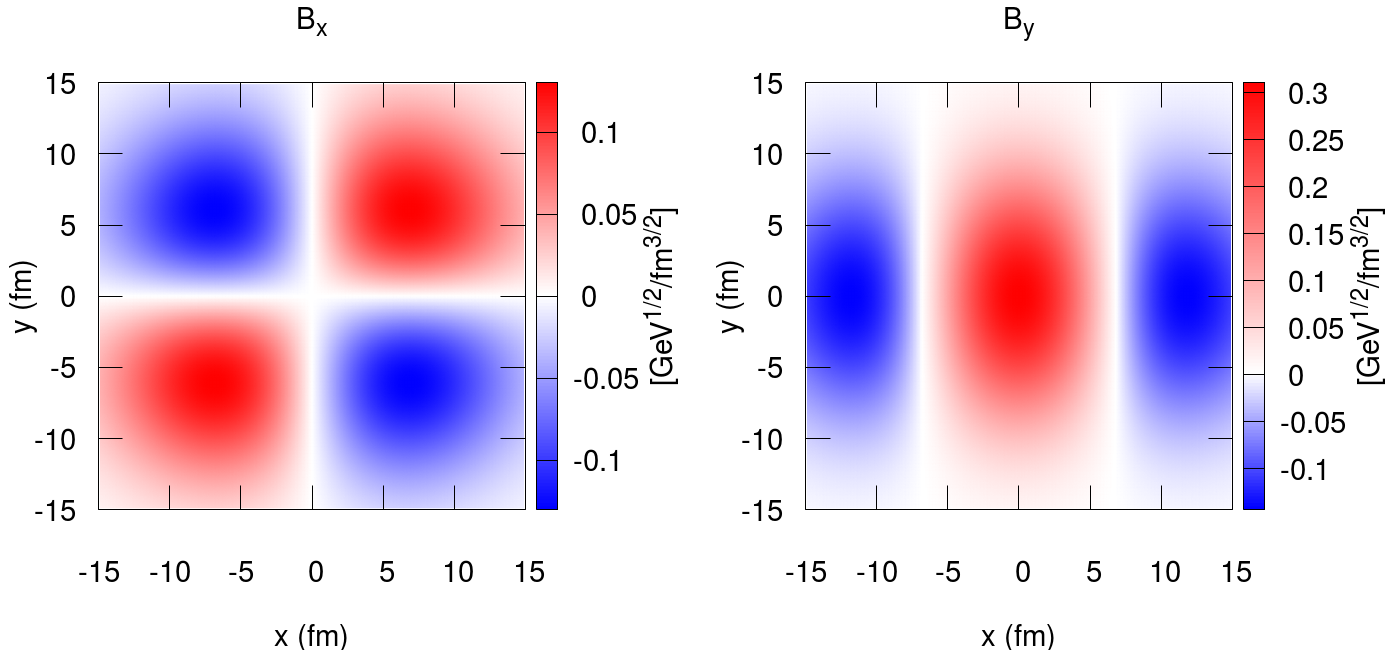}
	\caption{(color online) The initial spatial distribution of the components of the magnetic field $\vec{B}$, computed using the method described in Ref.~\cite{Tuchin:2013apa}. The parameters are for the reaction Au+Au, b=10 fm at $\sqrt{s_{NN}}$=200 GeV.}
	\label{initial_b}
\end{figure*}

\begin{figure}
	\center
	\includegraphics[width=0.45\textwidth]{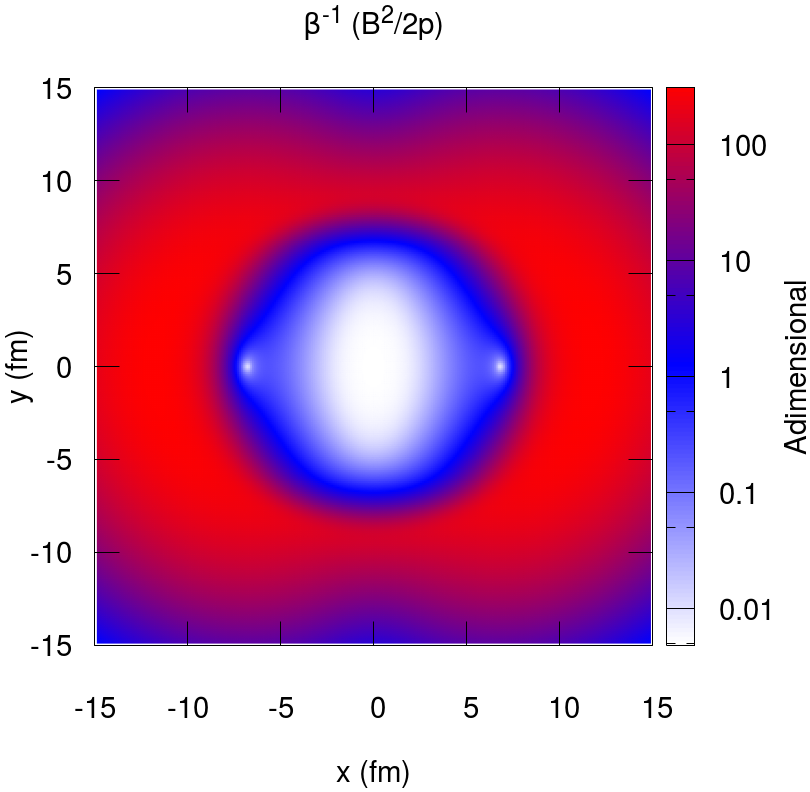}
	\caption{(color online) The initial ratio $1/\beta\!=\! B^2/2p$ between magnetic and thermal pressure in the transverse plane. The parameters are for the reaction Au+Au, b=10 fm at $\sqrt{s_{NN}}$=200 GeV.}
	\label{initial_beta}
\end{figure}

Our choices of the parameters for the initial conditions are summarized in Table (\ref{tabinpar}). The initial distribution of the thermal pressure, the magnetic field and the ratio of thermal to magnetic pressure in the transverse plane are shown in Figs. (\ref{initial_pressure}),(\ref{initial_b}) and (\ref{initial_beta}) for Au+Au, b=10 fm reactions at $\sqrt{s_{NN}}$=200 GeV.

We always use the same initial conditions for the initial energy density distribution, but for the initial magnetic field we consider two cases:
\begin{enumerate}
	\item $\vec{\textbf{B}}=0$ (no magnetic field)
	\item $\vec{\textbf{B}} \neq 0$ and $\sigma=5.8$ MeV
\end{enumerate}

In the first case we consider a pure hydrodynamical simulation, without magnetic field. In the second case we assume that in the pre-equilibrium phase there is a medium with finite constant electrical conductivity $\sigma=5.8$ MeV, which allows to compute an initial magnetic field distribution as in Ref.~\cite{Tuchin:2013apa}, shown in Fig. (\ref{initial_b}).

We assume that at the time $\tau_0$ the fluid is in local thermal equilibrium, its electrical conductivity $\sigma$ becomes infinite and that the magnetic field generated by the fast moving electric charges contained in the protons of the nuclei is converted into the magnetic field of the fluid, while, consistently with the hypothesis that initially the fluid is at rest and it has infinite electrical conductivity, we assume that there is no initial electric field in the fluid frame (otherwise, for Eq. (\ref{eq:ohm2}), we should have also initial non null fluid velocity). We neglect dissipative effects and we assume that the fluid obeys the $e\!=\!p/3$ EoS. 

We run the simulation until thermal freeze-out, when the energy density is below $150\,\textrm{MeV}/\textrm{fm}^3$. Then we compute the spectra and the elliptic flow of the pions produced. Here we adopt the Cooper-Frye prescription~\cite{DelZanna:2013eua,cooperfrye74}, without any modification to the distribution function due to the electromagnetic interaction.

\subsection{Results}

\begin{figure}
	\center
	\includegraphics[width=0.45\textwidth]{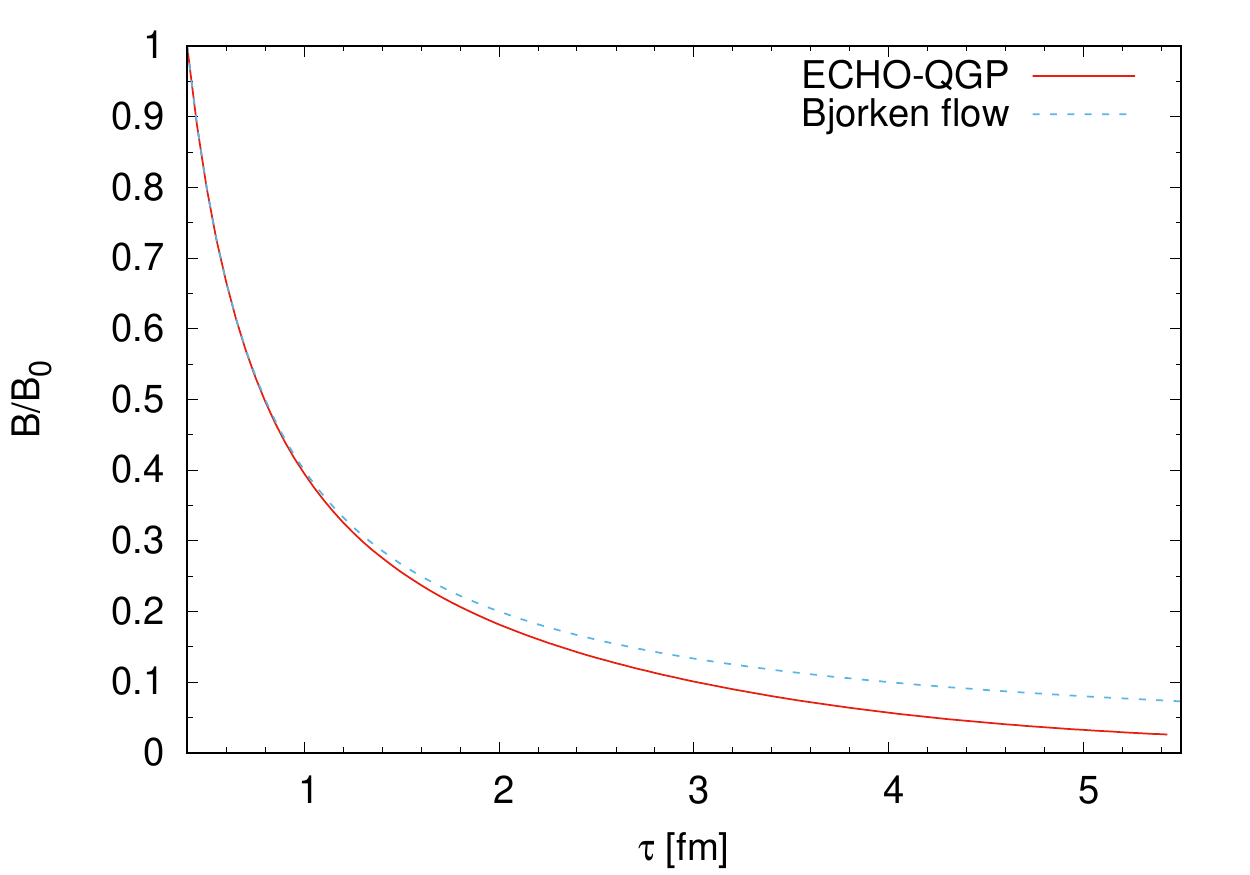}
	\caption{(color online) $B/B_0=\sqrt{B_iB^i(\tau)}/\sqrt{B_iB^i(\tau_0)}$, with $\tau_0=0.4$ fm/c. Comparison between the decay of the magnitude of B in the center of the grid during the 2D+1 RMHD evolution and the decay expected for a Bjorken flow, following the analytic law $\tau_0/\tau$. The parameters are for the reaction Au+Au, b=10 fm at $\sqrt{s_{NN}}$=200 GeV.}
	\label{B_decay}
\end{figure}

\begin{figure}
	\center
	\includegraphics[width=0.45\textwidth]{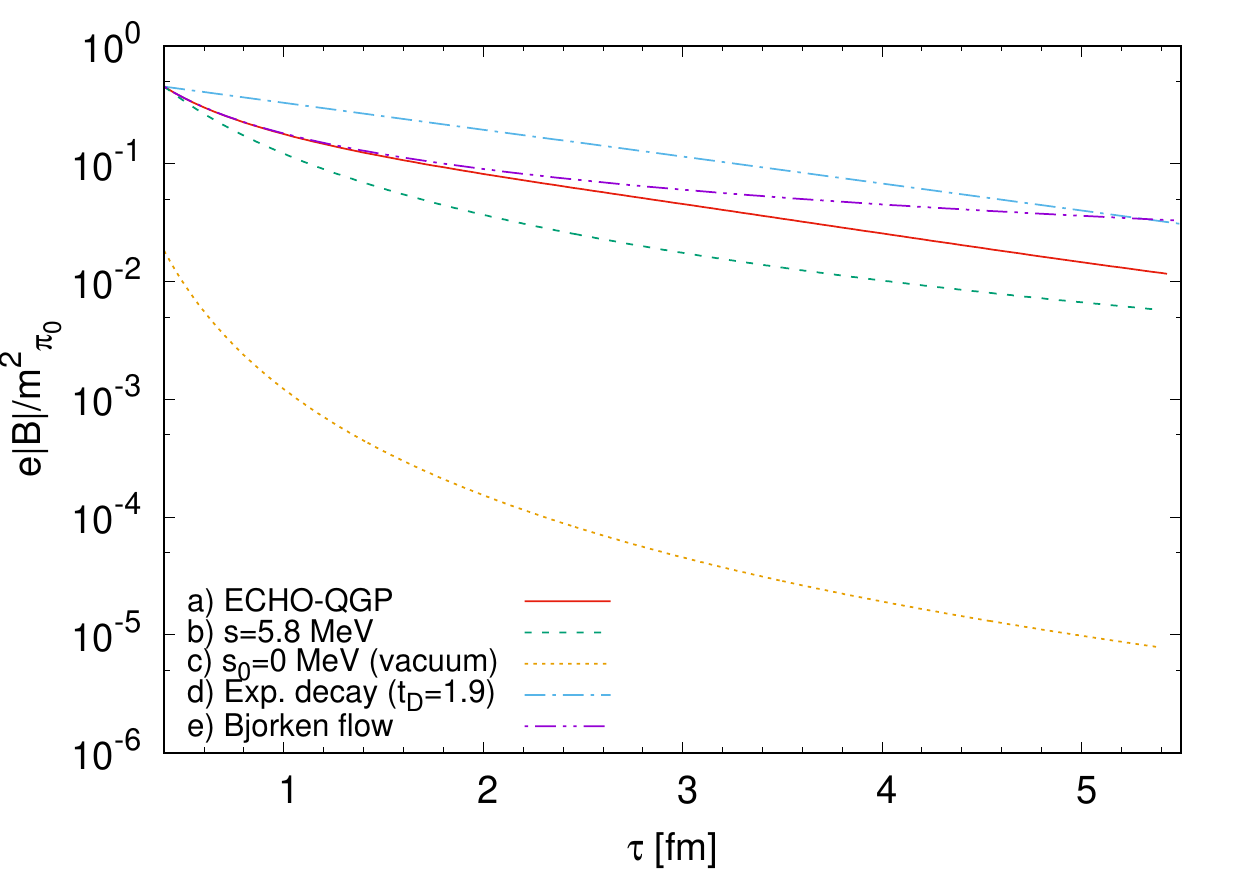}
	\caption{(color online) Comparison of the time evolution of the magnitude of the magnetic field (in neutral pion mass units squared) at the center of the grid in five different cases: a) with ECHO-QGP, as described in this section, computing the initial conditions assuming $\sigma=5.8$ MeV b) magnetic field generated by the electric charges of the two colliding nuclei moving in a medium with uniform and constant electrical conductivity $\sigma\!=\!5.8$ MeV, i.e. the same approach exploited to provide the initial conditions (explained in details in Ref.~\cite{Tuchin:2013apa}), but now adopted for the whole time interval c) same as in case b), but assuming zero electrical conductivity $\sigma\!=\!0$ MeV (vacuum) d) assuming an exponential decay of the magnetic field as modeled in Ref.~\cite{pang16}, with $t_D\!=\!1.9$ e) Bjorken flow. The parameters are for the reaction Au+Au, b=10 fm at $\sqrt{s_{NN}}$=200 GeV.}
	\label{B_decay_2}
\end{figure}

In  Fig. (\ref{B_decay}) and (\ref{B_decay_2}) we compare the decay of the magnetic field in the ideal 2D+1 RMHD simulation in the center of the of overlap region of the two nuclei (i.e. in the center of the grid: $x\!=\!y\!=\!z=\!\eta\!=0$) with some common analytical models. Fig. (\ref{B_decay}) shows the comparison between the decay of the magnitude of B in the center of the grid during the 2D+1 RMHD evolution and the decay expected for a Bjorken flow, following the analytic law $\tau_0/\tau$. Fig. (\ref{B_decay_2}) show the comparison of the time evolution of the magnitude of the magnetic field (in neutral pion mass units squared) at the center of the grid in five different cases:\begin{enumerate}[label=(\alph*)]
	\item ECHO-QGP 2D+1 RMHD evolution starting from initial conditions as described in this section, with $\sigma=5.8$ MeV
	\item time evolution of the magnetic field computed using the same approach exploited to provide the initial conditions (explained in details in Ref.~\cite{Tuchin:2013apa}), assuming assuming a medium with uniform and constant electrical conductivity $\sigma\!=\!5.8$ MeV 
	\item same as in case b), but assuming zero electrical conductivity $\sigma\!=\!0$ MeV (vacuum)
	\item exponential decay of magnetic field as modeled in Ref.~\cite{pang16}, with $t_D\!=\!1.9$
	\item Bjorken flow
\end{enumerate}  

\begin{figure}[!htb]
	\center
	\includegraphics[width=0.45\textwidth]{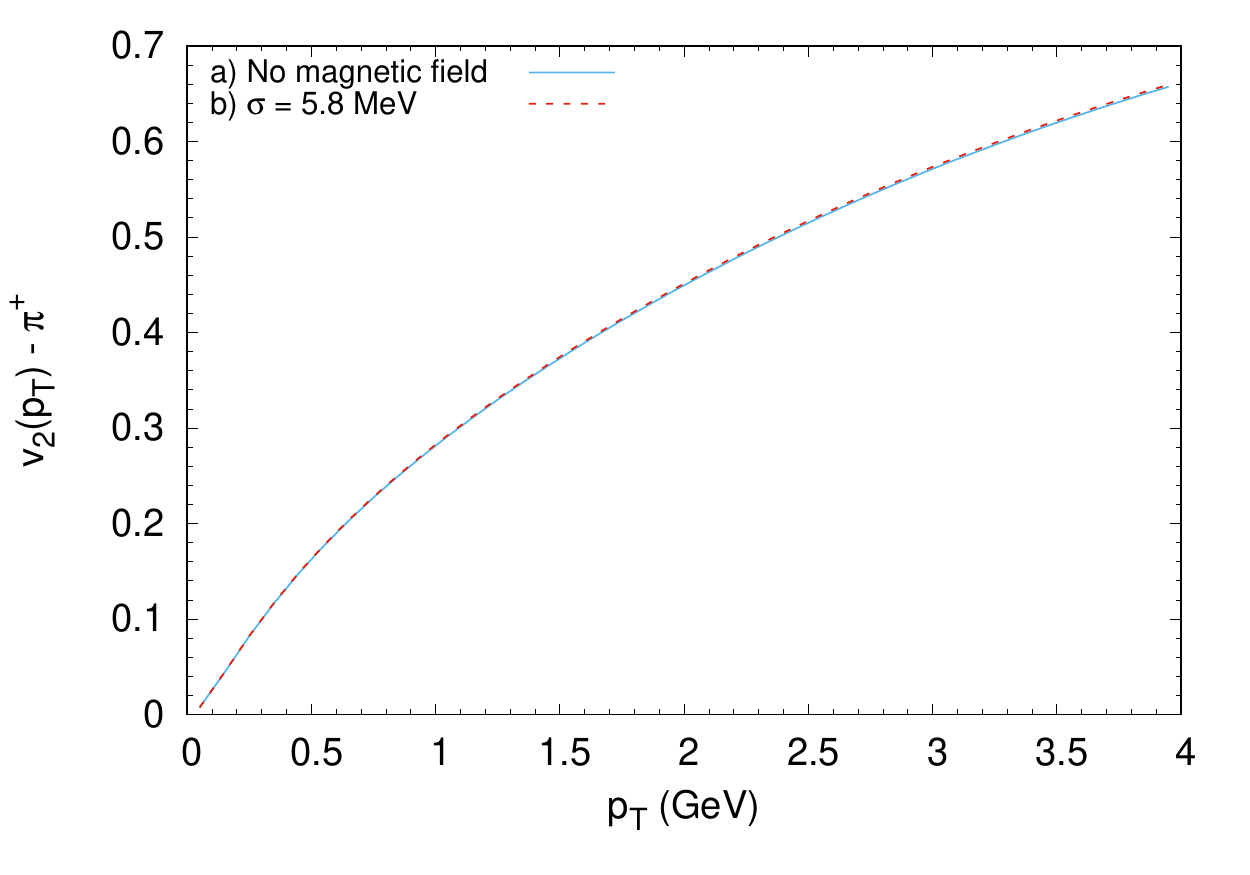}
	\caption{(color online) $v_2$ of $\pi^+$ in two cases: a) Without magnetic field, b) With an initial magnetic field computed assuming $\sigma=5.8$ MeV. The parameters are for the reaction Au+Au, b=10 fm at $\sqrt{s_{NN}}$=200 GeV.}
	
	\label{v2_rhic}
\end{figure}

\begin{figure}[!htb]
	\center
	\includegraphics[width=0.45\textwidth]{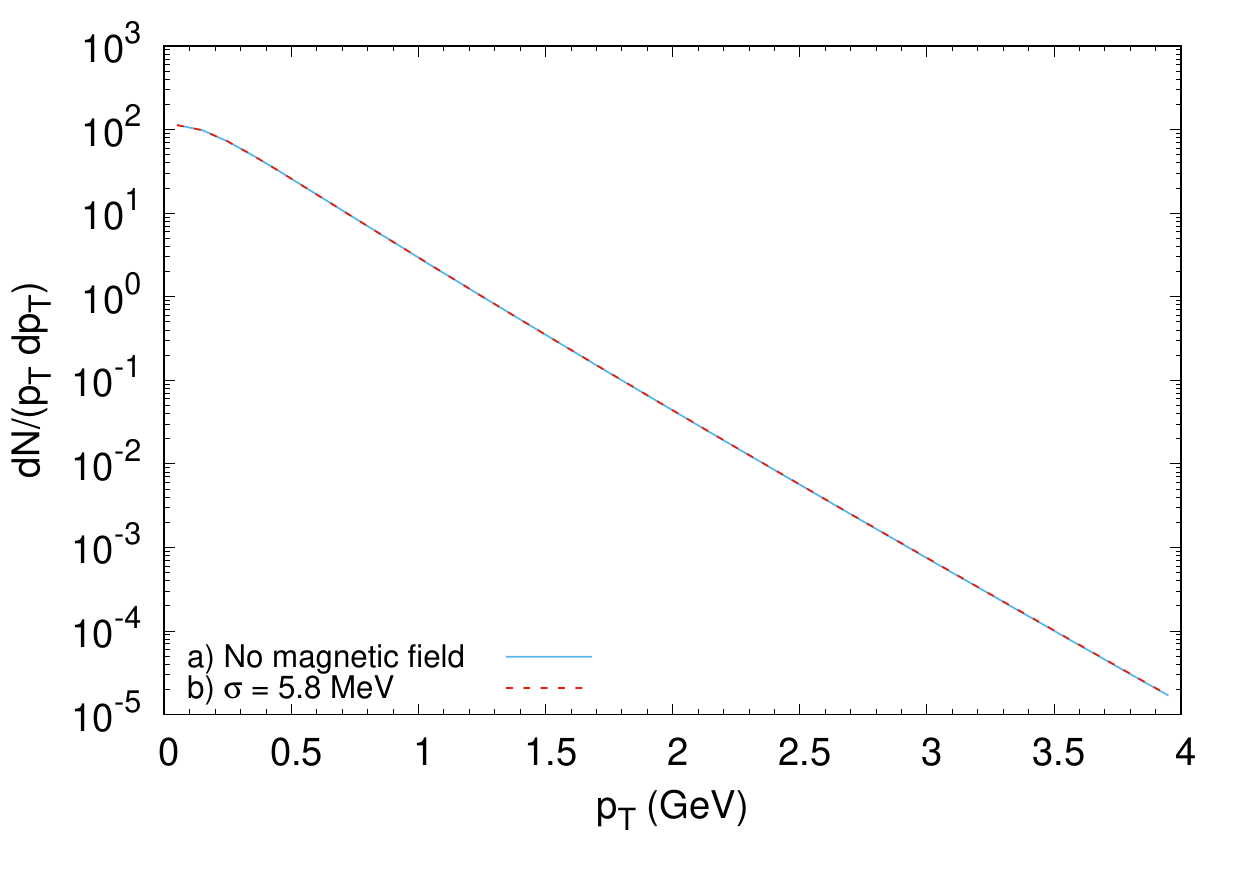}
	\caption{(color online) Transverse momentum distribution of $\pi^+$ in two cases: a) Without magnetic field, b) With an initial magnetic field computed assuming $\sigma=5.8$ MeV. The parameters are for the reaction Au+Au, b=10 fm at $\sqrt{s_{NN}}$=200 GeV.}
	
	\label{dNdpT_plot}
\end{figure}

We notice that the expansion of the fluid in the transverse plane leads to a faster decrease compared to the case of a pure longitudinal Bjorken-flow~\cite{roy15} and tends to become roughly exponential. However, the decay of the magnetic field of the fluid is still slower than in the case that the fields are generated by two electric charges moving in opposite directions in a uniform medium with constant finite electrical conductivity, as in Ref.~\cite{Tuchin:2013apa}, especially if there is no medium at all and the electric charge propagates in empty space. We stress that this comparison between different decay rates is based on a simplified model of HIC. In a 3D+1 simulation,  adopting a more realistic EoS and including dissipative effects, the decay rate of the B-field might be considerably quantitatively different. 

In Fig. (\ref{v2_rhic}) and (\ref{dNdpT_plot}) we compare the elliptic flow and the transverse momentum distribution of pions, computed with the Cooper-Frye prescription~\cite{DelZanna:2013eua,cooperfrye74}, with and without the presence of an initial magnetic field, computed as described in the previous section of this article. According to our current results, the presence of a magnetic field with a magnitude and spatial distribution evaluated according to Ref.~\cite{Tuchin:2013apa} seems to have a negligible impact both on the pion spectra and on the elliptic flow.  This is in contrast to Ref.~\cite{PhysRevD.93.054042} where was suggested that the magnetic field might substantially influence the anisotropic flow. In Ref.~\cite{moha11} it was indeed found that a significant enhancement of the elliptic flow might be possible. A direct comparison with our results is however not possible because of the many differences compared to our approach. However, in contrast to Ref.~\cite{PhysRevD.93.054042,moha11} and the present study, Ref.~\cite{pang16} reported the opposite result, namely a reduction of the anysotropic flow. This was attributed to the effects of the magnetic squeezing. However the model at Ref.~\cite{pang16} does not satisfy the divergence-free condition for the magnetic field. There the magnetic field has a rather large magnitude and it is not completely coupled with the fluid.

\section{Conclusions, discussion and outlook}\label{sec:conclusions}
We presented the extension of the ECHO-QGP code to the relativistic magnetohydrodynamic regime, in the limit of infinite electrical conductivity, i.e. without taking into account any resistive effect. In the present version, the code has been tested with an ideal-gas EoS, either in the presence of a finite mass-density or in the ultrarelativistic regime ($p=e/3$). After introducing the physics equations on which the code is based, we gave an overview of their numerical implementation. Then, we illustrated the results of several tests to validate the implementation. Since our final aim is to exploit the code to study the evolution of the Quark-Gluon Plasma formed in Heavy-Ion collisions, we showed first applications in this context, adopting simplified initial conditions.

Due to the (on average) small ratio of the magnetic to thermal pressure, the magnetic field does not seem to significantly affect the fluid evolution and we observed only a tiny effect on inclusive hadronic observables such as the elliptic flow and transverse momentum spectra of pions. However, in our approach the magnitude of the initial magnetic field could have been underestimated, possibly because in the pre-equilibrium phase we considered the electrical conductivity $\sigma$ as constant, while there are some evidences that it increases with the temperature~\cite{Aarts2015,PhysRevC.91.044903,PhysRevLett.111.172001,yin14}. Other authors, employing different initial conditions for the magnetic field, found a non-negligible effect of the latter on the hadron elliptic-flow~\cite{moha11,PhysRevD.93.054042,pang16}. Clearly this would affect the estimate of the viscosity-to-entropy $\eta/s$ ratio obtained by comparison of hydrodynamic results with experimental data: if, for example, part of the hadron $v_2$ in non-central collisions arose from the magnetic field, one should reduce the contribution from the hydrodynamic expansion, via for instance a larger value of $\eta/s$. 

Our preliminary results suggest also that the formation of a deconfined conductive plasma, compared to the case of the vacuum, might slow down the decay of the initial magnetic field generated by the colliding nuclei, possibly affecting non-perturbative phenomena relying on the presence of huge magnetic fields to show up. Since our study refers to the case of an ideal plasma, with infinite electrical conductivity, our results have to be considered as an upper limit on the lifetime of the magnetic field produced in heavy-ion collisions. 

However, the recent estimates both from lattice QCD computations~\cite{Aarts2015,PhysRevC.91.044903,PhysRevLett.111.172001} and fitting of experimental data~\cite{yin14} point toward high, but finite value for the electrical conductivity of the QGP. For a quantitative comparison with experimental data this has to be taken into account including the effects of the electrical resistivity. We expect a considerably acceleration of the decay of the magnitude of the magnetic field compared to our studies.

As a next step, we plan to evaluate better the role of the initial magnitude and spatial distribution of the magnetic fields, performing full 3D+1 simulations, already possible with the present setup. This will allow one to explore a broader range of possible initial conditions under different models~\cite{PhysRevC.92.064902}, using a more realistic EoS. 

The next development of the code will involve the inclusion of dissipative effects (shear and bulk viscosity and a finite electric conductivity), using the numerical techniques presented in \cite{DelZanna:2013eua} and \cite{DelZanna21082016}
and already implemented in previous versions of the ECHO code, going beyond the approximation of an ideal plasma. A major conceptual achievement would be represented by the inclusion in our setup of anomalous currents, allowing one to provide a consistent description of the CME and to estimate the possibility of disentangling it from other charge-separating effects related to the presence of strong electromagnetic fields.

Then, indeed, it would be necessary to modify the Cooper-Frye formula by taking into account the presence of an electromagnetic field and of a non uniform spatial distribution of electric charges. After that, for a proper comparison with experimental data, one should compute the effects on the final particle spectra and on collective flows, in the post-freeze-out phase, of decays, elastic collisions and of magnetic deflections by the Lorentz force.

Finally, we deem that applications of numerical calculations performed with the present relativistic MHD version of the ECHO-QGP code could be also relevant for cosmological (generation of the primordial magnetic fields~\cite{0034-4885-79-7-076901}) or astrophysical studies. For instance, the sudden transition from an hadronic to a QGP-like equation of state in a proto-magnetar (phase transition to a \emph{quark star}) has been recently suggested as a possible explanation for the observed cases of (long) Gamma-Ray Burst events with double prompt emission peaks~\cite{Pili16}.

\section*{Acknowledgments}
G. Inghirami thanks V. Roy, L. Rezzolla, L. Pang and M. D'Elia for fruitful discussions and useful suggestions. G. Inghirami was supported by a GSI grant in cooperation with the John von Neumann Institute for Computing. G. Inghirami also gratefully acknowledges support from the Helmholtz Research School on Quark Matter Studies and from Helmholtz Graduate School for Hadron and Ion Research. M. Haddadi Moghaddam would like to thank the ministry of science and technology of Iran and the Physics Department and the INFN section of Torino for warm hospitality and partial financial support during part of this work. The computational resources were provided by the INFN - Sezione di Firenze, by the Frankfurt Institute for Advanced Studies and by the Center for Scientific Computing (CSC) of the Goethe University. This work was supported by the University of Florence grant "Fisica dei plasmi relativistici: teoria e applicazioni moderne".
\appendix
\section{Appendix: Propagation of linear perturbations in the plasma}\label{app:perturbations}
In this appendix we want to present a study of the propagation of small perturbations in a relativistic plasma embedded in a constant magnetic field. Although this represents a standard MHD subject, we think it is useful for the reader to explicitly re-derive the main results for the case of an ultra-relativistic plasma addressed in this paper, with no conservation equation for the mass density, at variance with usual astrophysical studies.
We then perform small fluctuations around a homogeneous background, keeping in the equations only terms linear in the fluctuations. Taking into account that $\gamma\sim{\cal O}(\delta^2)$ one has (we consider the case of a one-dimensional flow along the $z$-axis)
\beq
\begin{aligned}
	&u^\mu=[1,0,0,\delta v], \quad p=p_0+\delta p,\\
	&e=e_0+\delta e, \quad b^\mu=b_0^\mu+\delta b^\mu.
\end{aligned}
\eeq
Notice that the index $0$ in the magnetic field is used to denote its unperturbed background value and not as a covariant index. Clearly, fluctuations in the pressure and energy density are related by the Equation of State. 
\subsection{Magnetosonic waves}\label{app:fast}
We firs want to evaluate the velocity of propagation of \emph{magnetosonic} disturbances. This will be relevant for the study of the self-similar one-dimensional flow described by the Lyutikov solution given in  Sec.~\ref{sec:Lyutikov}. We focus then on the propagation along the $z$-axis (i.e. $\delta=\delta(t,z)$) of the following perturbations
\beq
u^\mu=[1,0,0,\delta v]+{\cal O}(\delta^2),\quad
b^\mu=[0,b_0+\delta b,0,0]+{\cal O}(\delta^2)
\eeq
where, to linear order in the fluctuations, $B_0=b_0$ and $\delta B\approx \delta b$, so that one can identify the magnetic field in the laboratory and in the comoving frame. The system of RMHD equations reduces to
\begin{subequations}
	\begin{align}
	&\partial_t(\delta e)+b_0\partial_t(\delta b)+(e_0+p_0+b_0^2)\partial_z(\delta v)=0\\
	&\partial_z(\delta p)+b_0\partial_z(\delta b)+(e_0+p_0+b_0^2)\partial_t(\delta v)=0\\
	&\partial_t(\delta b)+b_0\partial_z(\delta v)=0
	\end{align}
\end{subequations}
Let us now perform a Fourier analysis of the fluctuations, inserting in the above the ansatz $\delta=\delta_{\omega,k}e^{-i\omega t+ikz}$. From the last equation, one gets for the magnetic field (turning out to fluctuate \emph{in phase} with the velocity)
\beq
\delta b_{\omega,k}=b_0(k/\omega)\delta v_{\omega,k},\label{eq:Bmagnetosonic}
\eeq
which can be substituted in the other two equations. Using an Equation of Ste of the kind $\delta p= c_s^2 \delta e$, one gets:
\begin{subequations}
	\begin{align}
	&\omega\, \delta e_{\omega,k}-(e_0+p_0)k\, \delta v_{\omega,k}=0\\
	&k\, c_s^2\,\delta e_{\omega,k}+[b_0^2(k^2/\omega)-(e_0+p_0+b_0^2)\omega]\,\delta v_{\omega,k}=0.
	\end{align}
\end{subequations}
The system has non-trivial solutions only if its determinant vanishes, i.e.
\beq
b_0^2k^2-(e_0+p_0+b_0^2)\omega^2+(e_0+p_0)c_s^2k^2=0,
\eeq
whose solution provides the dispersion relation $\omega=\omega(k)$
\beq
\omega^2=\frac{(e_0+p_0)c_s^2+b_0^2}{e_0+p_0+b_0^2}k^2\equiv c_f^2 k^2,
\eeq
which allows one to identify the \emph{fast magnetosonic speed} $c_f$. In the case of an ideal ultra-relativistic plasma $e_0=3p_0$ and $c_S^2=(1/3)$, so that one gets
\beq
{c_f^2=\frac{4p_0+3b_0^2}{3(4p_0+b_0^2)}}\,,\label{eq:fast1}
\eeq
which corresponds to the zero mass-density limit of Eq. (3) of the paper by Lyutikov and Hadden. 
In terms of the thermal to magnetic-pressure ratio 
\beq
{\cal B}\equiv\frac{p_0}{b_0^2/2}
\eeq
one gets
\beq
{c_f^2=\frac{2{\cal B}+3}{3(2{\cal B}+1)}}\,.\label{eq:fast2}
\eeq
\subsection{Alfv\'en waves}\label{app:Alfven}
Alfv\'en waves are MHD excitations which propagates along the lines of the unperturbed magnetic field. In full generality we will consider the evolution of the following perturbations (still neglecting ${\cal O}(\delta^2)$ terms in the fluctuations)
\beq
u^\mu\approx[1,0,0,\delta v],\quad b^\mu\approx[0,b_0+\delta b^x,\delta b^y, \delta b^z],
\eeq
where we take $\delta=\delta(t,\x_\perp)$: we will see that only the dependence on $x$, i.e. the direction of the unperturbed magnetic field, matters. We start considering the equations for the evolution of the components of the magnetic field.  To linear order in the fluctuations we have:
\beq
\partial_t\delta b^x\approx\partial_t\delta b^y\approx 0.
\eeq
If initially absent, no field perturbation develops along the $x$ and $y$ directions, perpendicular to the velocity fluctuation. Hence, in the following we set $\delta b^x= \delta b^y=0$.
On the other hand, from Faraday's law one has
\beq
\partial_t\delta b^z=b_0\partial_x\delta v,
\eeq
so that, employing the Fourier ansatz $\delta=\delta_{\omega,k}e^{-i\omega t+i k_x x+i k_y y}$ ($\k_\perp=(k_x,k_y)=(k\cos\theta,k\sin\theta)$), one gets
\beq
\delta b_{\omega,k}=-b_0(k\cos\theta/\omega)\delta v_{\omega,k}.\label{eq:BAlfven}
\eeq
The magnetic field develops a $z$-component, fluctuating \emph{in opposition of phase} with respect to the velocity. Let us now move to the equation for the energy and the fluid velocity. Notice that, to linear order, $\theta\approx\partial_x\delta v^x+\partial_y\delta v^y+\partial_z\delta v^z\approx 0$. Furthermore, since the fluctuations involve only the $z$-component of the B-field, one has $\partial_\mu b^2\approx 2b_0\partial_\mu\delta b^x\approx 0$. For the energy one gets then simply 
\beq
\partial_t\delta e\approx 0.
\eeq
For the Euler equation one gets instead:
\beq
(e_0+p_0+b_0^2)\partial_t\delta v^z-b_0\partial_x\delta b^z=0.
\eeq
In Fourier space one has then
\beq
(e_0+p_0+b_0^2)\,\omega\,\delta v_{\omega,k}+b_0\, k_x\,\delta b_{\omega,k}=0,
\eeq
which, employing Eq.~(\ref{eq:BAlfven}), leads to 
\beq
\omega^2\!=\!\frac{b_0^2}{e_0+p_0+b_0^2}k_x^2.
\eeq
The perturbation propagates then along the $x$-axis (the direction of the unperturbed magnetic field) with group velocity equal to the \emph{Alfv\'en speed} $v_g^x=({d\omega}/{dk_x})=v_A$, where
\beq
{v_A^2=\frac{b_0^2}{e_0+p_0+b_0^2}},
\eeq
which corresponds to the weak-fluctuation ($\eta\to 0$) limit of the exact result quoted in Eq.~(\ref{alfvenspeed}).
Assuming an ideal-gas EoS $e_0=3p_0$, the latter can be expressed in terms of ${\cal B}$ as
\beq
{v_A^2=\frac{1}{1+2{\cal B}}}\,,
\eeq
in agreement with Eq.~(3) of Lyutikov paper~\cite{PhysRevE.85.026401}, once setting to zero the contribution from the mass-density.
\clearpage

\end{document}